\documentclass[10pt]{article}	
\usepackage{float}		
\usepackage{times}		
\usepackage{graphicx}		
\usepackage{url}
\usepackage{lscape}
\usepackage{ulem}
\usepackage{epstopdf}
\usepackage{longtable}
\usepackage{amsmath}
\usepackage{amssymb}
\usepackage{multicol}
\usepackage[font={small}]{caption}
\usepackage{subcaption}
\usepackage{abstract}
\usepackage{fancyvrb}
\usepackage{verbatim}
\usepackage[procnames]{listings}
\usepackage[bottom]{footmisc}
\usepackage{tabto}
\usepackage{paralist} 
\usepackage{xcolor}
\usepackage[document]{ragged2e}
\usepackage{txfonts} 
\usepackage{microtype}
\usepackage{enumitem}  
\usepackage{comment} 
\usepackage[T1]{fontenc}
\usepackage{wrapfig}
\usepackage{tabularx}

\usepackage{placeins}
\usepackage{caption}
\usepackage[round]{natbib}
\usepackage{bm}
\usepackage{listings}
\usepackage{hyperref}

\newcommand{\aap}{A\&A}
\newcommand{\pasp}{PASP}
\newcommand{\apjl}{ApJL}

\newcommand{\apjs}{ApJS}

\hypersetup{
    colorlinks,
    linkcolor={blue!70!black},
    citecolor={blue!70!black},
    urlcolor={blue!70!black}
}
\newcommand{\farcs}{\ensuremath{.\!\!^{\prime\prime}}} 
\newcommand{\degr}{\hbox{\ensuremath{^{\circ}}}} 
\newcommand{\arcsec}{\ensuremath{^{\prime\prime}}}

\DeclareGraphicsExtensions{.pdf,.png,.jpeg}

\def\memonr{2023-L1-01}

\title{\vspace{-1.25cm} 
\textcolor{gray}{\textit{L1 Processing and Validation}}\\
\vspace{0.40cm}
\textbf{Metadata for the Flux Density Calibration of the April 2018 \\ 
Event Horizon Telescope Data}
} 

\author{ \renewcommand{\thefootnote}{\arabic{footnote}} 
J. Y. Koay$^{1}$, 
C. Romero-Ca\~nizales$^{1}$,
L. D. Matthews$^{2}$,
M. Janssen$^{3}$,
L. Blackburn$^{4,5}$,
R. P. J. Tilanus$^{6}$,\\
J. Park$^{7}$,
K. Asada$^{1}$,
S. Matsushita$^{1}$,
A.-K. Baczko$^{3}$,
N. La Bella$^{8}$,
C.-K. Chan$^{6,9,10}$,
G. B. Crew$^{2}$,\\
V. Fish$^{2}$,
N. Patel$^{4}$,
V. Ramakrishnan$^{11,12}$,
H. Rottmann$^{3}$,
J. Wagner$^{3}$,
K. Wiik$^{13}$,\\
P. Friberg$^{14}$, 
C. Goddi$^{8,15}$, 
S. Issaoun$^{4}$, 
G. Keating$^{4}$, 
J. Kim$^{6,16}$, 
T.P. Krichbaum$^{3}$,
D. Marrone$^{6}$, \\
G. Narayanan$^{17}$, 
A. Roy$^{3}$, 
I. Ru\'{\i}z$^{18}$, 
S. S\'anchez$^{18}$, 
P. Torne$^{18,3}$
J. Weintroub$^{4,5}$
}

\def\affiliations{
\hspace{0.5cm}$^{1}$\textit{Academia Sinica Institute of Astronomy and Astrophysics, No.1, Sec. 4, Roosevelt Rd, Taipei 10617, Taiwan, R.O.C.} \\
\hspace{0.5cm}$^{2}$\textit{Massachusetts Institute of Technology Haystack Observatory, 99 Millstone Road, Westford, MA 01886, USA} \\
\hspace{0.5cm}$^{3}$\textit{Max Planck Institut f\"ur Radioastronomie (MPIfR), Auf dem H\"ugel 69, 53121 Bonn, Germany} \\
\hspace{0.5cm}$^{4}$\textit{Center for Astrophysics $|$ Harvard \& Smithsonian, 60 Garden St., Cambridge, MA 02138, USA} \\
\hspace{0.5cm}$^{5}$\textit{Black Hole Initiative at Harvard University, 20 Garden Street, Cambridge, MA 02138, USA}\\
\hspace{0.5cm}$^{6}$\textit{Steward Observatory and Department of Astronomy, University of Arizona, 933 N. Cherry Avenue, Tucson, AZ 85721, USA}\\
\hspace{0.5cm}$^{7}$\textit{Department of Astronomy and Space Science, Kyung Hee University, 1732, Deogyeong-daero, Giheung-gu, Yongin-si, Gyeonggi-do 17104, Republic of Korea}\\
\hspace{0.5cm}$^{8}$\textit{Department of Astrophysics / IMAPP, Radboud University, P.O. Box 9010, 6500 GL Nijmegen, The Netherlands}\\
\hspace{0.5cm}$^{9}$\textit{Data Science Institute, University of Arizona, 1230 N. Cherry Ave., Tucson, AZ 85721}\\
\hspace{0.5cm}$^{10}$\textit{Program in Applied Mathematics, University of Arizona, 617 N. Santa Rita, Tucson, AZ 85721}\\
\hspace{0.5cm}$^{11}$\textit{Finnish Centre for Astronomy with ESO, FI-20014 University of Turku, Finland}\\
\hspace{0.5cm}$^{12}$\textit{Aalto University Mets\"ahovi Radio Observatory, Mets\"ahovintie 114, FI-02540 Kylm\"al\"a, Finland}\\
\hspace{0.5cm}$^{13}$\textit{Tuorla Observatory, Department of Physics and Astronomy, University of Turku, Finland}\\
\hspace{0.5cm}$^{14}$\textit{East Asian Observatory, 660 N. A’ohoku Place, Hilo, HI 96720, USA}\\
\hspace{0.5cm}$^{15}$\textit{INAF - Osservatorio Astronomico di Cagliari, Via della Scienza 5, 09047, Selargius, CA, Italy}\\
\hspace{0.5cm}$^{16}$\textit{California Institute of Technology, 1200 East California Boulevard, Pasadena, CA 91125, USA}\\
\hspace{0.5cm}$^{17}$\textit{Department of Astronomy, University of Massachusetts, 01003, Amherst, MA, USA}\\
\hspace{0.5cm}$^{18}$\textit{Institut de Radioastronomie Millim\'{e}trique, Avenida Divina Pastora 7, Local 20, E-18012, Granada, Spain}
} 

\date{Dec 05, 2023 -- Version 1.1}
\def\memohistory{
\item {Dec 15, 2021 -- Document created}
}
\newif\ifshowhistory  

\begin{document} 

\begin{figure}[!t]
\begin{minipage}[t]{0.49\textwidth}
    \vspace{-0.13\linewidth}{\includegraphics[width=0.6\linewidth]{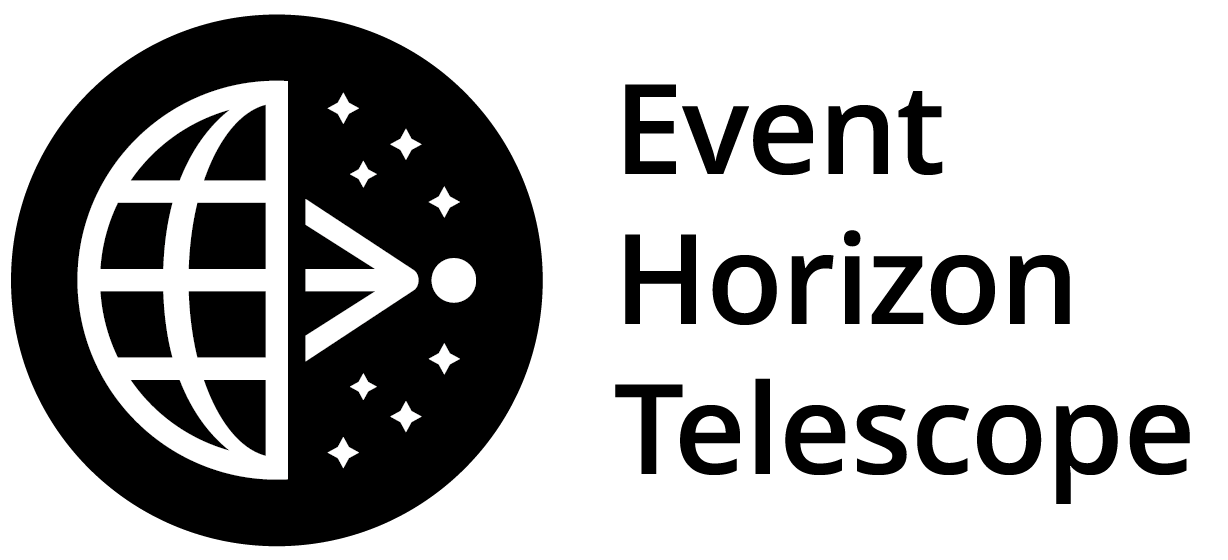}}
    \label{fig:my_label}
\end{minipage}
\begin{minipage}[t]{0.49\textwidth}
\begin{flushright} \large
\textbf{Event Horizon Telescope \\ Memo Series} 
\end{flushright}
\end{minipage}
\end{figure}

\begin{center}
EHT Memo \memonr{}
\end{center}

{\let\newpage\relax\maketitle}  
\thispagestyle{empty}		

\begin{flushleft}
{\small
\affiliations{}
}
\end{flushleft}

\ifshowhistory{
\setitemize{noitemsep,topsep=0pt,parsep=0pt,partopsep=0pt}  
\begin{flushleft}
\vspace{0.25cm}
Memo history:
\begin{itemize}
\itemsep0em
\memohistory{}
\end{itemize}
\end{flushleft}
}\fi


\begin{abstract} 
 The Event Horizon Telescope (EHT) observations carried out in 2018 April at 1.3\,mm wavelengths included 9 stations in the array, comprising 7 single-dish telescopes and 2 phased arrays. The metadata package for the 2018 EHT observing campaign contains calibration tables required for the a-priori amplitude calibration of the 2018 April visibility data. This memo is the official documentation accompanying the release of the 2018 EHT metadata package, providing an overview of the contents of the package. We describe how telescope sensitivities, gain curves and other relevant parameters for each station in the EHT array were collected, processed, and validated to produce the calibration tables.

\end{abstract} 

\newpage

\tableofcontents


\setlength{\parskip}{\baselineskip}
\parindent 0pt
\vspace{1.0cm}

\section{Introduction}\label{sec:introduction}

The Event Horizon Telescope (EHT) 2018 campaign was carried out over six tracks observed on 2018 April 21, 22, 24, 25, 27 and 28. There were a total of nine participating stations, of which two are phased arrays (ALMA and SMA) and seven are single dish stations (APEX, IRAM 30\,m, SPT, LMT, SMT, JCMT and GLT). The Greenland Telescope \citep[GLT;][]{inoueetal14,chenetal23} is a new addition to the 2018 EHT array. The EHT array also observed at double the bandwidth than in 2017. Four frequency bands were used, each with bandwidths of 1.875\,GHz for ALMA and 2.0\,GHz at the other stations. These bands were centered at frequencies of 213.1 (band 1), 215.1 (band 2), 227.1 (band 3), and 229.1\,GHz (band 4). Due to the mismatched sampling rates between ALMA and the other stations, a portion of each bandwidth is lost during correlation, resulting in correlated data of 1.856\,GHz per band (comprising 32 spectral windows of 58\,MHz each). Note that the GLT only observed in bands 3 and 4 in 2018. With the exception of ALMA, which observes in linear polarization, all the other EHT stations observed in circular polarizations. Visibility data on ALMA baselines are converted from mixed to circular polarization basis via the {\fontfamily{qcr}\selectfont PolConvert} software package \citep{martividaletal16} after correlation. Additionally, the JCMT observed only a single polarization (right circular polarization, RCP) throughout the 2018 campaign. In 2017, the JCMT switched between RCP and left circular polarization (LCP) observations on different tracks.

The processing of the EHT data from 2018 onwards is now separated into Level 1 (L1) and Level 2 (L2) tasks, each producing corresponding data packages. The L2 data package is still being defined at the time of writing. The L1 data package mainly comprises validated correlation products, and an associated metadata package containing information on telescope sensitivities and other parameters necessary for the flux density calibration of the correlated visibilities. The L1 data products are released to the EHT collaboration and the respective project principal investigators (PIs), including PIs external to the EHT collaboration, for further calibration and for archiving purposes. Within the EHT, the additional calibration steps required to produce science-ready data are referred to as the L2 processing stage; these L2 calibrated data are internal EHT collaboration products.

Prior to 2018, each telescope site provided the metadata required for the flux density calibration in different formats and with varying levels of refinement \citep[for details, see][]{janssenetal19, issaounetal17a}. Processing the metadata to generate calibration tables (in the standard, widely used ANTAB\footnote{\url{http://www.aips.nrao.edu/cgi-bin/ZXHLP2.PL?ANTAB}} format) was thus time-consuming and labor-intensive. For 2018, steps were taken to standardize both the format and contents of the requested metadata packages across all stations, where possible, as well as the scripted pipeline reading and processing these packages. This allows for a streamlined approach that is easy to maintain and reuse for future EHT epochs. The long term goal is for the metadata collection to be automated as much as possible, and incorporated into the VLBI Monitor\footnote{\url{https://vlbimon1.science.ru.nl/}} system. 

\subsection{Scope of this Memo}\label{sec:scope}

In this memo, we describe the collection, processing and validation of the 2018 metadata necessary for the absolute flux-density calibration of the 2018 EHT observations, and the contents of the metadata products released as part of the L1 package. In Section~\ref{sec:concepts}, we give a brief overview of the basic concepts for the flux calibration of EHT data. We then provide details on the raw metadata collection process for 2018 in Section~\ref{sec:rawcollect}. Section~\ref{sec:dpfugains} describes the determination of the DPFUs and gain curves for each station, while Section~\ref{sec:processing} describes additional processing and validation of the system noise temperatures and phased array SEFDs, as required for the generation of the ANTAB calibration tables. Section~\ref{sec:l1products} summarizes the contents of the L1 metadata package, as well as a brief summary of the metadata properties. Telescope acronyms and codes used in this memo are found in Appendix~\ref{app:ids}. The codes used to reference the different observing tracks, and details on the participating stations during each track, are also provided in Appendix~\ref{app:ids}.

\section{Basic Concepts for EHT Flux Calibration}\label{sec:concepts}

We define here the basic terms and concepts for EHT and (sub-)mm VLBI flux calibration, which will be used throughout this memo. The EHT memos by \citet{issaounetal17a} and \citet{janssenetal19} for the calibration of the EHT 2017 data provide more detailed descriptions of these concepts.

To convert the visibility amplitudes from correlation coefficients to units of flux density, taking into account the different telescope sensitivities across the array, the system equivalent flux density (SEFD) of each telescope needs to be determined. The SEFD is given by \citep[e.g.,][]{ehtpaperIII19}:
\begin{equation}\label{eqSEFD}
{\rm SEFD} = \dfrac{1}{\eta_{\rm ph}}\dfrac{T_{\rm sys}^{*}}{{\rm DPFU} \times g_E}~,
\end{equation}
where the degrees per flux density unit, DPFU, is a conversion factor from units of temperature (K) to units of flux density (Jy), and $g_E$ is the normalized elevation-dependent gain curve of the telescope. $\eta_{\rm ph}$ is the time-dependent phasing efficiency for a phased array, and has a value of unity for single dish telescopes. The DPFU can be estimated as:
\begin{equation}\label{eqDPFU}
{\rm DPFU} = \dfrac{\eta_{\rm A} A_{\rm geom}}{2k}~,
\end{equation}
where $\eta_{\rm A}$ is the aperture efficiency of the antenna, $A_{\rm geom}$ is the geometrical area of the antenna dish in units of m$^{2}$, and $k = 1.38 \times 10^{3}$ is the Boltzmann constant in units of Jy\,m$^2$/K.

$T_{\rm sys}^{*}$ is the effective system noise temperature, corrected for atmospheric attenuation, given by:
\begin{equation}\label{eqTsys}
T_{\rm sys}^{*} = \frac{e^{\tau}}{\eta_{\rm l}}T_{\rm sys}~, 
\end{equation}
where $\tau$ is the line-of-sight atmospheric opacity, and $\eta_{\rm l}$ is the forward efficiency \citep[implicitly assumed to be the same as the feed efficiency, accounting for rearward efficiency loss due to ohmic losses, spillover and scattering; see][for instance]{kutnerulich81}. It is important to note that $T_{\rm sys}^{*}$ differs from the standard $T_{\rm sys}$ value used in cm-wavelength VLBI, which does not account for atmospheric attenuation. Single dish (sub-)mm telescopes typically measure $T_{\rm sys}^{*}$ directly by placing a hot load of known temperature in the signal path. This is done regularly during or in between target scans to account for the time and elevation (airmass) dependence of $T_{\rm sys}^{*}$.

The gain curve of each station is parameterized as a function of elevation, $E$, following the equation \citep{janssenetal19}:
\begin{equation}\label{eq:gc}
g_E = 1 - B (E - E_0)^2 \,.
\end{equation}
This equation is fit to measurements of antenna temperature or antenna efficiency (normalized by their mean values at the peak of the curve) as a function of elevation at each station, to derive the best fit values of the parameters $B$ and $E_0$, both of which are coefficients. These two parameters are then converted to the polynomial gain coefficients used in the standard ANTAB calibration table format, given by:
\begin{equation}\label{eq:gc_coeff}
a_{\rm 0} = 1 - B{E_0}^2 \, , \quad a_{\rm 1} = 2B{E_0} \, , \quad a_{\rm 2} = -B\,.
\end{equation}
The coefficients beyond the second order are set to 0, following the parameterization of Equation~\ref{eq:gc} which, based on measurements in 2017 considers that a second order polynomial is sufficient to fit the gain curve of existing EHT stations. Note that there is still an uncertainty about the shape of the LMT gain curve (especially considering the dish has an active surface\footnote{\url{http://lmtgtm.org/telescope/telescope-description/}} to correct for gravitationally induced deformations), due to a lack of measurements (see Section~\ref{sec:LM}), especially since the completion of its upgrade to a 50\,m diameter antenna in 2017 December.

\section{Collection and Contents of the Raw Metadata Packages}\label{sec:rawcollect}

In 2020 October, a plan was put in place to compile the 2018 EHT metadata according to specific standards. The aim was to streamline the tasks during and after metadata collection, including metadata validation and the production of ANTAB calibration tables. After receiving feedback from site representatives during 2020 November, the checklist of required metadata and submission formats was finalized, and the majority of telescope representatives submitted their metadata packages in 2021 January. Cross-checks and validation of the raw metadata packages were conducted throughout 2021. Any technical and/or formatting issues discovered were reported back to the telescope representatives, so that the issues could be fixed and new versions submitted.

The raw metadata from each station comprise the following files:
\begin{itemize}
    \item \textbf{Effective system temperature ($T_{\rm sys}^{*}$) table} per observed track, with associated timestamps, VEX scan number and name of observed source for each scan. For validation purposes, additional columns include elevation, azimuth, opacity at zenith, effective temperature of the atmosphere ($T_{\rm atm}$) and ambient temperature in the receiver cabin ($T_{\rm amb}$). For single-dish stations, there is one $T_{\rm sys}^{*}$ column per polarization (RCP, LCP), per observing band (lower sidebands, i.e., band 1 \& band 2; and upper sidebands, i.e., band 3 \& band 4); this gives eight columns with $T_{\rm sys}^{*}$ values per scan for single-dish stations. For the SMA, RCP and LCP $T_{\rm sys}^{*}$ values are provided for each of the 7 antennas in the array, giving a total of 14 columns with $T_{\rm sys}^{*}$ values. For ALMA, phased $T_{\rm sys}^{*}$ values (and corrected for the phasing efficiencies) are provided in the tables. Since the bandpass information is also used when estimating the phased $T_{\rm sys}^{*}$, entries are provided for each of the 32 spectral windows in 32 columns. Therefore, for ALMA, one $T_{\rm sys}^{*}$ file is provided per band and per observing track, to avoid putting $T_{\rm sys}^{*}$ for all four bands in a single file with 32$\times$4 columns.
    
    \item \textbf{Flag table} per observed track, denoting the status and viability of each observed scan, with comments from the telescope operators on possible technical or weather issues. Each scan is assigned a flag code, which can be used to assist downstream processing in determining the data quality and in the excision of bad data. The flag codes stand for the following possible scenarios: success/no problems (S), not observed/unusable data (N), partial observation or late on source (P), not good or uncertain data quality (U), missing $T_{\rm sys}^{*}$ (T). More than one code can be applied to each scan record.
    
    \item \textbf{Phasing efficiency table} per observed track, for phased arrays only. For the SMA, the phasing efficiencies are provided per band and per polarization, for a total of 8 columns in each file. For ALMA, the values contained in the phasing efficiency tables are in linear polarization, and are not the values used to produce the final calibration tables, so are solely for diagnostic checks. The ALMA phasing efficiencies are also provided per band and per (linear, X and Y) polarization. 
    \item \textbf{Antenna efficiency and gain table} for single dish stations (with associated elevation and timestamps if available), by which the station DPFUs and elevation dependent gain curves can be derived.  The stations also had the option to directly provide DPFUs and gain parameters that they had estimated themselves.
\end{itemize}

The format of these tables was devised such that the header of each file and its name unequivocally identify it with a given observing track and antenna site. Examples of the table filenames, formats and contents are provided in Appendix~\ref{app:rawformat}. Each row of $T_{\rm sys}^{*}$ values and their corresponding flag code should ideally be associated with valid VEX scan numbers and source names. We acknowledge that in some stations the local software from where measurements are obtained, has certain restrictions such as the number of characters allowed for source names. Thus, sometimes source names will be shorter versions of the VEX names, which is dealt with by the processing pipeline (Section~\ref{sec:processing}).

When available, the telescope representatives also provided supporting data and documents, including antenna temperature measurements, observer logs, and descriptions of how the main quantities in the metadata were measured.

\section{Station Sensitivities and Gain Curves}\label{sec:dpfugains}

The DPFUs and gain curve coefficients of each station in the array are provided in the ANTAB table headers, which are then used by standard calibration software to perform absolute flux density calibrations on visibility data. A summary of the station DPFUs and gain curve coefficients for the EHT array in 2018 is provided in Table~\ref{dpfugainstable}, together with their estimated uncertainties. The DPFU uncertainties are expected to be the dominant source of systematic errors in the flux density calibration of EHT data, which includes uncertainties in the planet models used to derive their flux densities and between planet models used in different calibration software (typically 5 to 10\%). $T_{\rm sys}^{*}$ values are not expected to vary much within the typical scan duration of a few minutes, so their uncertainties are expected to be no larger than a few percent. The GLT is an exception (details in Section~\ref{sec:GL}), due to sub-optimal performance in 2018 when it was still only partially commissioned, where $T_{\rm sys}^{*}$ uncertainties are also significant. The following subsections describe how the DPFUs, gain curves, and their corresponding uncertainties were determined for each station. 

\begin{table*}[ht]
\caption{EHT station DPFUs and gain curve parameters. DPFU values are quoted for both lower (LSB) and upper (USB) sidebands and both polarizations (RCP and LCP). For phased arrays (ALMA and SMA), the DPFUs represent the combined sensitivity of all phased dishes. The parameters listed in the table are defined in Section~\ref{sec:concepts}. }\label{dpfugainstable}

\begin{tabularx}{\linewidth}{@{\extracolsep{\fill}}l c c c c c c c}
\hline\hline       
Station &  \multicolumn{5}{c}{DPFU (K/Jy)} & \multicolumn{2}{c}{$g_E$} \\ 
\cline{2-6}\cline{7-8}
& LSB-RCP & LSB-LCP & USB-RCP & USB-LCP & uncertainties & $B$ & $E_0$\\ 
\hline  
 ALMA\textsuperscript{a} & 1.19 & 1.19 & 1.19 & 1.19 & 
 $\pm$ 10\% & 0 & 0\\
 APEX\textsuperscript{b} & 0.0253 & 0.0262 & 0.0259 & 0.0270 & $\pm$ 5.5\% & 0.00002 $\pm$ 3.6\% & 36.6 $\pm$ 1.0\%\\
 GLT\textsuperscript{c} & - &  - & 0.00885 & 0.00885 & $\pm$ 7\% & 0 & 0\\
 LMT & 0.188 & 0.126 & 0.188 & 0.126 & $\pm$ 22\% & 0 & 0\\
 SMT & 0.0188 & 0.0188 & 0.0182 & 0.0182 & $\pm$ 6\% & 0.000082 $\pm$ 10.4\,\% & 57.6 $\pm$ 2.0\%\\
 JCMT\textsuperscript{d} & 0.0296 & - & 0.0296 & - & $\pm$(11--14)\% & 0 & 0\\
 IRAM 30\,m & 0.144 & 0.138 & 0.142 & 0.137 & $\pm$ 19\% & 0.00018 $\pm$ 5.3\% & 43.7 $\pm$ 1.3\%\\
 SMA\textsuperscript{e} & 0.040 & 0.040 & 0.040 & 0.040 & $\pm$(5 -- 15)\% & 0 & 0\\
 SPT & 0.00698 & 0.00731 & 0.00698 & 0.00731 & $\pm$ 10\% & 0 & 0\\
\hline
\end{tabularx}
\newline
\textsuperscript{a}\footnotesize{The ALMA DPFU uncertainty is based on the overall 10\,\% systematic uncertainties associated with the QA2 flux calibration tables.}\newline
\textsuperscript{b}\footnotesize{APEX measures DPFUs for each of the 4 bands separately. LSB and USB DPFUs shown here are the mean of the band 1 and 2 DPFUS, and band 3 and 4 DPFUs, respectively.}\newline
\textsuperscript{c}\footnotesize{GLT DPFU uncertainties are based on the scatter of 3 measurements, inclusive of flux density model errors, but may be larger due to significant antenna astigmatism. There is also a $\sim 15\%\, T_{\rm sys}^{*}$ uncertainty contributing to the total SEFD uncertainty.}\newline
\textsuperscript{d}\footnotesize{For the JCMT DPFU, the night-time value is shown.}\newline
\textsuperscript{e}\footnotesize{The SMA DPFU uncertainty is based on the dominant 5-15\,\% uncertainty on the phasing efficiency.}\newline
\end{table*}

\subsection{ALMA}\label{sec:AA}

The ALMA flux calibration parameters were determined by the ALMA Level 2 Quality Assurance (QA2) team for the EHT observations, who directly produce and submit the calibration tables in ANTAB format. The ALMA SEFDs are determined based on self-calibration gains and measured $T_{\rm sys}^{*}$ converted to circular polarizations via {\fontfamily{qcr}\selectfont PolConvert} \citep{martividaletal16}. Just as for 2017, an average DPFU value of 0.031 K/Jy corresponding to that of a single 12\,m dish is factored out of the SEFD estimates, and used in the ANTAB table header. Variations in the sensitivity of the phased ALMA array due to the total number of observing antennas and phasing losses are incorporated into the time-dependent system temperature values in the EHT ANTAB tables. A flat gain curve is assumed, since any elevation and time-dependent antenna gains are solved indirectly through self-calibration. Details on the ALMA QA2 calibration for 2017 are presented by \citet{goddietal19} and are also discussed briefly by \citet{janssenetal19} and \citet{ehtpaperIII19}. 

\subsection{APEX}\label{sec:AX}

The APEX DPFUs are determined by the APEX team from observations of Mars and Jupiter, based on planet flux densities derived using the {\fontfamily{qcr}\selectfont ASTRO} program within the IRAM {\fontfamily{qcr}\selectfont GILDAS}\footnote{\url{https://www.iram.fr/IRAMFR/GILDAS}} software package. The given uncertainties of 5.5\% include both the measurement errors and the 5\% quoted flux density uncertainties of the planet models. The determination of APEX calibration parameters is described in detail in the accompanying documentation provided with the raw APEX metadata package, contained within in the L1 metadata package (see Section~\ref{sec:l1products}). 

Although no gain dependence on elevation was found based on the 2018 planet antenna temperature measurements, quasar data from 2017 showed a curve with small gains (with no more than 5\% deviations from a flat gain curve), which was used for the 2017 EHT data calibration. We use the same gain curve for 2018, with parameters shown in Table~\ref{dpfugainstable}.

\subsection{GLT}\label{sec:GL}

The GLT was only partially commissioned during the 2018 observations, resulting in the telescope performance being sub-optimal at the time. A systematic, saddle-shaped deformation in the GLT antenna dish imprinted when the dish was lifted onto the support cone, resulted in significant astigmatism and a low measured aperture efficiency of $21.6 \pm0.9\%$ at 230\,GHz for the upper sidebands. As noted, the GLT did not observe in bands 1 and 2 in 2018. Due to RCP output instabilities in the efficiency measurement data, only the LCP values could be used; the LCP aperture efficiency is applied to both polarizations. This aperture efficiency was derived based on scans of Venus, using the flux density model from the {\fontfamily{qcr}\selectfont CASA} data reduction software \citep{butler12}, and taking into account the extended size of the planet and the beam size of the GLT. With a geometric dish diameter of 12\,m, this translates into a DPFU value of $0.0088 \pm 0.0004$\,K/Jy, considering only the measurement errors. Adding the quoted 5\% uncertainties of the Venus flux density model \citep{butler12} in quadrature gives a 7\% total uncertainty. Note that these uncertainties are derived based only on the three available efficiency measurements in March 2018, before the dish surface improvements were carried out. The uncertainties are thereby likely to be larger and are difficult to quantify, due to additional systematic errors arising from the dish astigmatism. Nevertheless, due to instabilities in the output of the continuum detector used for the $T_{\rm sys}^{*}$ measurements in 2018, the $T_{\rm sys}^{*}$ measurements are estimated have uncertainties of up to $\sim 15\%$ , thus dominating the total SEFD uncertainties. Detailed descriptions of these issues, and the full derivation of the GLT calibration parameters are described in a separate memo \citep{koayetal23}.

Due to the extreme Northern latitude location of the telescope, observed sources typically do not vary significantly in elevation throughout an observing track. Therefore a flat gain curve is assumed.

\subsection{LMT}\label{sec:LM}

Due to an emergency situation, LMT observations were cancelled midway through the 2018 EHT campaign, so observers were unable to measure the aperture efficiency after the campaign as planned. LMT aperture efficiencies were subsequently measured in 2020 January, and are provided by the telescope team in the gain tables. We use these 2020 measurements to be representative of the 2018 DPFU values, since no work was done on the antenna between the two dates. 

\begin{figure}[ht]
    \centering
    \includegraphics[width=0.8\textwidth]{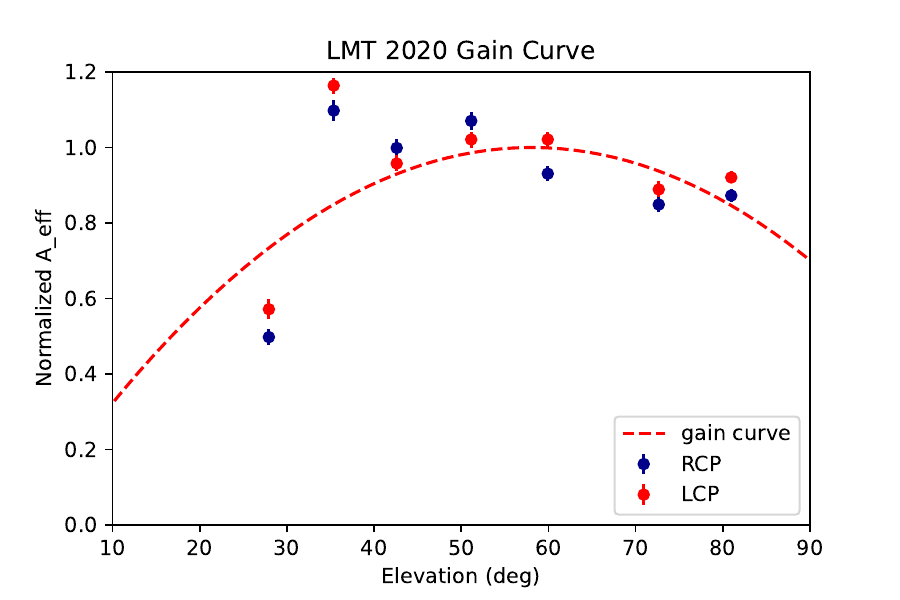}
    \caption{The LMT aperture efficiencies in 2020, normalized by the mean aperture efficiency at the gain curve plateau (at elevations of 40-60\degr{}) for each polarization. The dashed line shows the best fit gain curve using the normalized aperture efficiencies at both polarizations.}\label{fig:LMTgc}
\end{figure}

The LMT aperture efficiencies between 28-80\degr{} are shown in Figure~\ref{fig:LMTgc}, normalized over the mean of the aperture efficiencies at the peak/plateau range of elevations, i.e., between 40\,\degr{} to 60\degr{}, assuming that the measurement at 35\degr{} is an outlier. These aperture efficiencies were determined based on observations of R~Leo. Assuming a 50\,m diameter dish, we used the mean aperture efficiencies of 26.5\% in RCP and 17.7\% in LCP to derive a DPFU of 0.188\,K/Jy in RCP and 0.126\,K/Jy in LCP.

The gain curve cannot be reliably characterized due to insufficient data points at low elevations. In particular, there is an uncertainty regarding how the active surface panels, designed to compensate for elevation-dependent deformations of the dish, actually affect the gain curve. We therefore tentatively assume a flat gain curve for the LMT with large uncertainties in the DPFU, up to 22\%, considering the 1$\sigma$ standard deviation of the measured antenna efficiencies. Additional fixes may be needed downstream to improve the LMT calibration, at least until the gain curve can be characterized once the telescope resumes operations, which have stopped since March 2020 due to the Covid-19 pandemic.

\subsection{SMT} \label{sec:MG}

Measurements of the SMT aperture efficiency and gain curves for the EHT 2017 observations are described in detail in the memo by \citet{issaounetal17b}. Since no additional work was carried out at the telescope between 2017 and 2018, the gain curve used for the 2017 observations is retained for 2018. The aperture efficiency data were reprocessed by the SMT team, using the Butler 2018 planet models of Jupiter and Mars (Bryan Butler, private communication), giving a beam size of 34\farcs8 and an aperture efficiency of $0.66 \pm 0.04$ for the lower sidebands (both polarizations), and $0.64 \pm 0.04$ for the upper sidebands (both polarizations). These translate into DPFUs of $0.0188 \pm 0.011$\,K/Jy for bands 1 and 2, and $0.0182 \pm 0.011$\,K/Jy for the upper sidebands, consistent with the 2017 values. We note that these DPFU values were provided directly by the SMT team in the ANTAB header format, together with the gain curve polynomial coefficients. 

\subsection{JCMT} \label{sec:MM}

\begin{figure}[ht]
    \centering
    \includegraphics[width=0.7\textwidth]{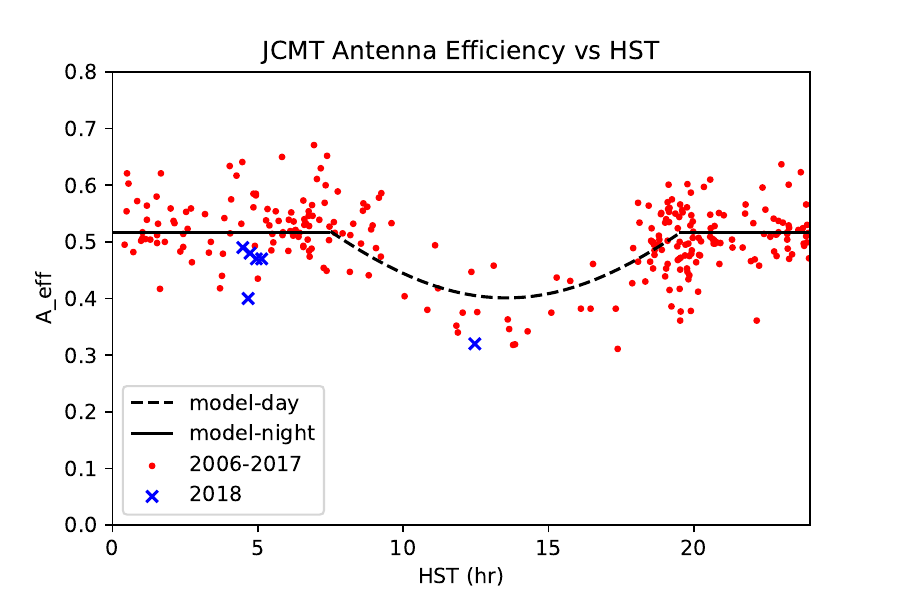}
    \caption{The JCMT antenna efficiencies measured between 2006 and 2017 (red points) and that measured in 2018 (blue crosses). The dashed line shows flat gains during night-time HST, and the dip in antenna efficiencies during the day time.}\label{fig:JCMTgc}
\end{figure}

Long-term characterization of the JCMT antenna reveals no significant elevation-dependent gains, so a flat elevation-gain curve is used. However, there is a known time-dependence of the gains (Figure~\ref{fig:JCMTgc}), determined from antenna aperture efficiency measurements between 2006 and 2017 \citep{issaounetal18}. The aperture efficiencies remain flat during night-time (1930 to 0730 Hawaii Standard Time, HST), and then dip during the day. These time variations in the aperture efficiencies are caused by deformations in the dish due to thermal gradients induced by solar heating during the day. For the 2018 observations, we use the same piece-wise model-fit of the gains determined from the 2006 to 2017 data by \citet{issaounetal18}, assuming that the gains behavior has not changed in 2018. The 2018 antenna efficiency measurements obtained in the daytime indeed show consistent trends (Figure~\ref{fig:JCMTgc}). However, we normalize the night-time gains ($g_{t}$) to unity, giving:
\begin{align}\label{eq:JCMTgc} 
{g_{t}} & = 1.0 & {\rm for\,1930\,to\,0730\,HST} \\ \nonumber
{g_{t}} & = 1.938 - 1.161 \times {\rm exp}\left( - \dfrac{({\rm HST} - 13.550)^{2}}{167.701} \right) & {\rm for\,0730\,to\,1930\,HST}
\end{align}
In the ANTAB calibration tables, these time-dependent gains are accounted for by dividing the $T_{\rm sys}^{*}$ values of each JCMT scan by the corresponding gains value. 

The six measurements of Mars and Uranus obtained by the JCMT in 2018 show a systematic decrease in the aperture efficiencies compared to the mean of the 2006-2017 values (Figure~\ref{fig:JCMTgc}). Using only the mean night-time aperture efficiency value of 46.2\% in 2018, provided in the JCMT gain tables, we estimate the 2018 night-time JCMT DPFU to be 0.0296\,K/Jy in RCP (in 2018, the JCMT observed only in RCP). While the aperture efficiencies are measured at 221\,GHz, we assume this value applies to all four bands. Due to the lack of measurements in 2018, we assume DPFU uncertainties of $\pm 11\%$ in the night-time, and $\pm 14\%$ during the day, similar to that between 2006-2017, and consistent with the scatter in the 2018 aperture efficiency measurements.

\subsection{IRAM 30\,m} \label{sec:PV}

The IRAM 30\,m telescope team provided DPFUs with associated uncertainties for the upper and lower sidebands and for both polarizations, which they determined from Mars and 3C279 observations in 2018. The two calibrators give systematically offset DPFU values, but are still consistent within the 1$\sigma$ errors, so they are deemed insignificant. For the ANTAB tables, we use the mean DPFU of these two calibrators for each band and polarization, as given in Table~\ref{dpfugainstable}.

\begin{figure}[ht]
    \centering
    \includegraphics[width=0.8\textwidth]{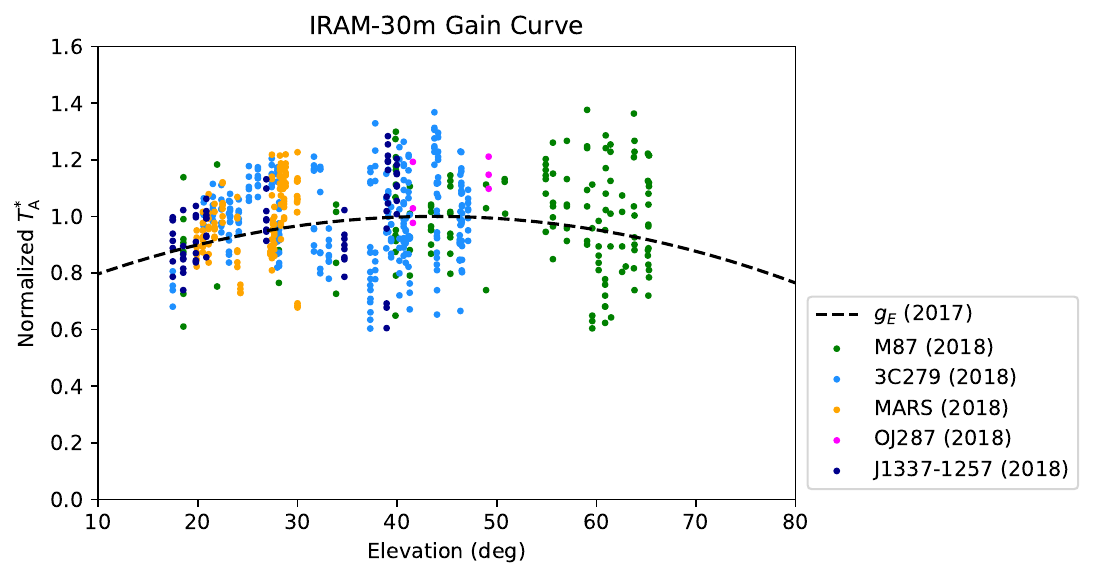}
    \caption{The IRAM 30\,m telescope antenna temperatures ($T_{\rm A}^{*}$) measured in 2018, normalized for each source by their mean values within the elevation range of 25 - 50\degr{}. Only $T_{\rm A}^{*}$ values obtained when $\tau < 0.5$ and for measured FWHM beam sizes less than $15\arcsec$ are used. Large $T_{\rm A}^{*}$ outliers have also been flagged. The black dashed line shows the 2017 gain curve.}\label{fig:PVgc}
\end{figure}

Due to bad weather during the 2018 observations, the antenna temperature measurements show a large scatter and have large uncertainties, and are thus not ideal for the estimation of the gain curve. Only antenna temperature measurements from M87 observations are available at high elevations (Figure~\ref{fig:PVgc}), and the scatter is large. Additionally, some sources including Mars, are observed over a limited range of elevations, making it difficult to normalize the antenna temperatures consistently for all sources. In Figure~\ref{fig:PVgc} for example, the normalization is done using the mean antenna temperature values of each source within the elevation range of 25 - 50\degr{}, which is not ideal. We therefore re-use the 2017 IRAM 30\,m gain curve \citep[][]{janssenetal19} for the 2018 observations, since they are not expected to vary significantly.

\subsection{SMA}\label{sec:SW}

The SMA DPFUs are provided by the SMA team for each observing track, antenna, band and polarization. They are estimated based on observations of Callisto for the April 21, 22, 24 and 25 tracks, and of Ganymede for the April 27 track. Due to poor weather, no good calibration solutions were available to derive the DPFU for the April 28 track, so we use the April 27 DPFU values for Apr 28, as recommended by the SMA team. The phased DPFU for all 7 antennas observing in 2018 is 0.040\,K/Jy, for a single-dish aperture efficiency of $\sim$75\%.  

For the ANTAB calibration tables, the DPFU of the SMA is set to 1.0. The actual measured DPFUs are used to modify the $T_{\rm sys}^{*}$ measurements for each antenna, before the $T_{\rm sys}^{*}$ values (in units of Jy) are added together in phase \citep[see Equation 3 in][]{janssenetal19}. More details are provided in Section~\ref{sec:process_phased}.

The SMA has a flat gain curve as a function of elevation \citep{matsushitaetal06}. This is because each of the array antennas adjusts its subreflector focus position as a function of elevation in real-time, based on the measured elevation-dependent focus curves. The focus curves are different from antenna to antenna, but a common focus curve (i.e., averaged over all antennas) is used for each of the antennas to avoid introducing any path length differences, and thus delays, between antennas. 

\subsection{SPT} \label{sec:SZ}

Four measurements of the SPT aperture efficiency were provided by the station representatives, determined from observations of Saturn and Jupiter on 2018 April 28 at elevations of 16 to 22\degr{}. The 2018 flux density models for Saturn and Jupiter at 221.1\,GHz (Bryan Butler, private communication) were used to characterize the SPT beam shape. The mean aperture efficiencies are 24.5$\pm$0.8\% in RCP and 25.7$\pm$0.8\% in LCP. Assuming a 10\,m diameter dish, this gives DPFU values of 0.0069$\pm$0.0002\,K/Jy and 0.0073$\pm$0.0002\,K/Jy in RCP and LCP, respectively. With uncertainties of up to 10\% in the flux density model of Saturn, adding this in quadrature to the measurement uncertainties gives 10\% uncertainties in total for the SPT DPFUs. The total error budget is thus dominated by the flux density model uncertainty. We assume that these DPFU values are consistent across both the lower and upper sidebands. The low aperture efficiencies are a result of the telescope being under-illuminated, operating at sensitivities closer to that of a 6\,m dish. 

As for the 2017 calibration of SPT, we ignore the elevation-dependent gains of the SPT, since sources typically do not vary much in elevation throughout the observing track due to its geographical location.

\section{Processing and Validation of System Temperatures and Phased Array SEFDs}\label{sec:processing}

\subsection{Processing Pipeline}\label{sec:pipeline}

We built a {\fontfamily{qcr}\selectfont python} pipeline (available on a private GitHub repository\footnote{\url{https://github.com/cristinaroca/EHTmd}. Send your GitHub ID to \url{cromero@asiaa.sinica.edu.tw} to receive an invitation.}) to process the metadata from both the single-dish and phased array stations, while performing a number of validation routines. An independent pipeline\footnote{\url{https://github.com/mpifr-vlbi/l1-calibration}. Send your GitHub ID to \url{mjanssen@mpifr-bonn.mpg.de} to receive an invitation.} has also been developed for processing of the EHT metadata in the future. 
Calibration tables are then generated for the whole EHT array in the standard ANTAB format used in many radio astronomical data reduction software packages. 

After defining some global parameters pertaining to an EHT run, the pipeline automatically draws information from VEX files and performs a cross-match with the metadata tables submitted by each station. Hence, only some code sections (written specifically for antennas needing special processing in a particular observing run) would need to be re-written or commented out when running the pipeline for different observing runs (see Section~\ref{sec:process_single}).

\subsection{Single Dish Stations}\label{sec:process_single}

The pipeline cross-matches available VEX files with the files submitted by each station (see Section \ref{sec:pipeline}). This helps verify that every expected file in the metadata packages actually exists, as well as to identify erroneous information associated with metadata measurements. During verification, the pipeline issues warnings and provides both lists of missing files and lists of affected scans. Relevant metadata files automatically open for inspection in real time so that a clarification request from telescope representatives is made promptly, and/or solutions are implemented when needed.

While an issue is resolved, the user can decide on-the-fly whether the affected track/antenna/scan rows should be flagged or if the issue can be safely ignored. If flagging the data, the options are to delete the row completely, or to have their $T_{\rm sys}^{*}$ values reset to NaN ({\it not a number} in computing notation) and later replaced by modelled values. 

Here we describe the main processing and validation steps applied to single dish stations (including issues detected and how these were resolved).

\subsubsection{System Noise Temperature Timestamps}\label{sec:sd_tshift}

We applied a shift to $T_{\rm sys}^{*}$ timestamps to correspond to the middle of their respective scans. We do this only for stations providing a single $T_{\rm sys}^{*}$ measurement per scan, i.e., all single-dish stations observing in 2018. This eases the a-priori flux calibration process downstream, considering that different stations use different strategies for observing $T_{\rm sys}^{*}$; some stations measure $T_{\rm sys}^{*}$ before the target scan, while others measure after, or sometimes a mixture of both. Standard data processing software like {\fontfamily{qcr}\selectfont AIPS} \citep{greisen03} requires that the $T_{\rm sys}^{*}$ timestamp be between the start and stop time of the scan. Typically, a time `offset' parameter in the ANTAB table is used to shift all the $T_{\rm sys}^{*}$ timestamps of a station by a constant value in time to move them in between the appropriate scans. However, due to different sources having different observing scan lengths, as well as inconsistencies in the time gaps between the actual $T_{\rm sys}^{*}$ measurements and the scan start/stop times for a particular station, applying a single, constant time offset value to each station is insufficient for adjusting all the $T_{\rm sys}^{*}$ timestamps in the observing track correctly in between the station's corresponding scan start/stop times. 

\subsubsection{Metadata Clean-up and Flagging}\label{sec:sd_flag}

When comparing metadata and VEX files, the pipeline identifies source name discrepancies and a warning is raised. In the 2018 metadata we found seven such discrepancies. Five of these corresponded to the use of alternative VEX source names for which we directly implemented source name replacements in the pipeline. The additional discrepancies were due to scans reflecting $T_{\rm sys}^{*}$ information for a calibrator (OMC1 in the case of the SMT, and Saturn in the case of the LMT) used for pointing instead of information for the target source (OJ287 and SgrA*, respectively). After inspection of the affected metadata file, we flagged the associated $T_{\rm sys}^{*}$ measurements to be later replaced by modelled values.

Some of the early metadata submissions also showed scan number discrepancies when compared to the corresponding VEX files. Telescope representatives were informed and requested to submit revised metadata packages.

Additional series of checks were aimed at finding contiguous scans on different sources sharing the same timestamps and/or $T_{\rm sys}^{*}$ values, due to e.g., errors in data entry or during the production of the raw metadata package at each station. A couple of LMT scans had such a problem, and after careful inspection, we deemed these values as unreliable and decided to flag them for replacement.

Using information from the VEX files, the pipeline obtains global variables and calculates expected antenna elevations per scan by taking into account the telescope location and source position in the sky at the time of observing. Doing this is useful in two ways: 
\begin{itemize}
    \item We can compare calculated elevations with those provided in the raw metadata packages and obtain a list of track/antenna/scan entries where discrepancies are larger than 5\degr{}. For the LMT, we found seven scans in track e18c21 (corresponding to April 21, see table~\ref{tab:tracks}) with elevation discrepancies of 8\degr{} up to 28\degr{} and we noticed that the largest difference between $T_{\rm sys}^{*}$ and VEX timestamps in these scans was shorter than 20 minutes. Looking at the trend of elevation and  $T_{\rm sys}^{*}$ variation in time for neighbouring scans on the same source, we determined that these entries should have their $T_{\rm sys}^{*}$ flagged for replacement. In addition, a total of 235 SMA $T_{\rm sys}^{*}$ measurements in different tracks had their original elevation values differing from the calculated ones by about $8\degr{}$ up to $65\degr{}$. Most of these corresponded to values measured during the long gaps between scans of target sources. After verifying that there were other measurements for the same scans closer in time to the actual VEX scan times, these entries were flagged to be completely deleted. 
    \item Some scans were observed by a particular station, but do not have any associated $T_{\rm sys}^{*}$ nor elevation entry in the system noise temperature table. This includes three LMT scans having an ``S'' flag code in the flag table. We thus assigned a calculated elevation to such entries. In addition, we obtained an opacity value by interpolating $\tau$ from neighbouring scans, to accompany the new elevation information. 
\end{itemize}
Having reliable elevation values for scans whose original elevation values were dubious, as well as for scans with no elevation information, allows us to obtain substitute model $T_{\rm sys}^{*}$ values for them (see Section \ref{sec:sd_model}).

Finally, the pipeline attempts to automatically detect outliers. For this, we implement an initial elevation and opacity curve fitting (see details in section \ref{sec:sd_model}), and deviations from the modelled curves are obtained. To ensure outlier candidates are narrowed down automatically as much as possible, we also used the modified z--score as an additional outlier indicator. This method uses the median absolute deviation, which is more robust against outliers than the median or the mean. Once an outlier is identified, its associated $T_{\rm sys}^{*}$ is flagged. This section of the code for automated flagging requires further improvement with better flagging algorithms. 

In addition, interactive plots are generated to aid the identification of outlier candidates (see Section \ref{ssec:val_plots}). Since all the values are tagged by track/antenna/scan, we can easily identify the origin of suspicious values and inspect them in their corresponding metadata files. 

The outlier candidates identified by the code and then visually inspected, mostly occur when there are rapid $T_{\rm sys}^{*}$ changes. This is the case in many SMT scans (see details in section \ref{sec:sd_model}). After inspection, we did not deem any of the candidates found in SMT nor any other antenna as true outliers.

Detailed information on the scans affected by elevation discrepancies and other issues described above, are noted in the auxiliary files generated by the pipeline, included in the metadata package (Section~\ref{sec:l1products}). 

\subsubsection{Additional Processing for Specific Antennas}\label{sec:sd_individual}

{\it LMT -- } Band and polarization labels have been corrected. After interaction with telescope representatives, it was determined that the $T_{\rm sys}^{*}$ values provided under the column headers {\fontfamily{qcr}\selectfont Tsys\_b1r}, {\fontfamily{qcr}\selectfont Tsys\_b1l}, {\fontfamily{qcr}\selectfont Tsys\_b2r}, and {\fontfamily{qcr}\selectfont Tsys\_b2l} in the {\fontfamily{qcr}\selectfont .tsys} tables (an example of this table is provided in Appendix~\ref{app:rawformat}) from the LMT raw metadata package  corresponded instead to upper and lower sideband in LCP, and to lower and upper sideband in RCP. 

{\it JCMT -- } The JCMT $T_{\rm sys}^{*}$ values are measured at the intermediate frequency (IF) of 4\,GHz, corresponding to bands 2 and 3 in Double-sideband (DSB) mode. We have assigned the same $T_{\rm sys}^{*}$ values to the other two bands, assuming they are comparable across the bands. Note that this may lead to larger uncertainties in the $T_{\rm sys}^{*}$ values in bands 3 and 4. The DSB $T_{\rm sys}^{*}$ values are then converted to single-sideband (SSB) equivalent values, using the empirically measured\footnote{\url{https://www.eaobservatory.org//jcmt/wp-content/uploads/sites/2/2019/05/RxA3m-SB-Notes-2018.pdf}} sideband ratio of 0.9. This is done by multiplying a factor of 1.9 to the $T_{\rm sys}^{*}$ of bands 3 and 4 (USB), and a factor of 2.11 to the $T_{\rm sys}^{*}$ in bands 1 and 2 (the LSB corresponding to the image sidebands). We then divided the $T_{\rm sys}^{*}$ values by the gains in night-time and day-time, as shown in equation \ref{eq:JCMTgc}.

\subsubsection{Modelling of System Noise Temperatures}\label{sec:sd_model}

After identifying scans affected by different issues, some $T_{\rm sys}^{*}$ entries will now appear as NaN and need to be replaced with alternative values.

We model $T_{\rm sys}^{*}$ as a function of elevation ($E$) and opacity ($\tau$). We first consider Equation~\ref{eqTsys} with $\tau = \tau_0/{\rm \sin}{(E)}$, where $\tau_0$ is the opacity at zenith. We note that at some stations, the zenith opacities are measured by a radiometer operating at a slightly different frequency from that of the actual EHT observing frequencies. However, the differences in $\tau$ are not expected to be significant. 

Taking into account the receiver temperature ($T_{\rm rx}$), the temperature that an antenna should measure when pointing at the empty sky, at a given elevation is given by:
\[
T_{\rm sys} = T_{\rm rx} +  T_{\rm atm}\eta_{\rm l}(1-{\rm e}^{-\tau}) + T_{\rm amb}(1-\eta_{\rm l}),
\]
where the contribution from the Cosmic Microwave Background radiation is regarded as negligible. Substituting $T_{\rm sys}$ in Equation~\ref{eqTsys}, we have \citep[][]{rohlfsandwilson06}:
\[
    T_{\rm sys}^{*} = \frac{{\rm e}^{\tau}}{\eta_{\rm l}} \left( T_{\rm rx} + T_{\rm atm}\eta_{\rm l}(1-{\rm e}^{-\tau}) + T_{\rm amb}(1-\eta_{\rm l}) \right)
\] 
\[
    \implies  
     T_{\rm sys}^{*} = \frac{{\rm e}^{\tau}}{\eta_{\rm l}}\left( T_{\rm rx} + T_{\rm atm}\eta_{\rm l} + T_{\rm amb} (1-\eta_{\rm l})\right) - T_{\rm atm} \,.
\]  
By considering ~ $Q_0 = \left( T_{\rm rx} + T_{\rm atm}\eta_{\rm l} + T_{\rm amb} (1-\eta_{\rm l})\right)\eta_{\rm l}^{-1}$ ~ and ~ $Q_1 = - T_{\rm atm}$, ~ we have:   
\begin{equation}\label{eq:tsysmod} 
     T_{\rm sys}^{*} = {\rm e}^{\tau_0/{\rm \sin}{(E)}}Q_0 + Q_1,
\end{equation}
which is the function we have used in the pipeline to find $T_{\rm sys}^{*}$ outliers, and to model missing  $T_{\rm sys}^{*}$ values. This approach alleviates the problem of not having  $T_{\rm atm}$ and $T_{\rm amb}$ values (they were not provided by some stations) as well as other quantities. We note that this equation is valid when $T_{\rm atm} = T_{\rm amb}$ and also in the cases when $T_{\rm atm} \neq T_{\rm amb}$ (e.g., for the GLT and the SPT).

\begin{figure*}[!ht]
\centering
\includegraphics[width=0.90\textwidth]{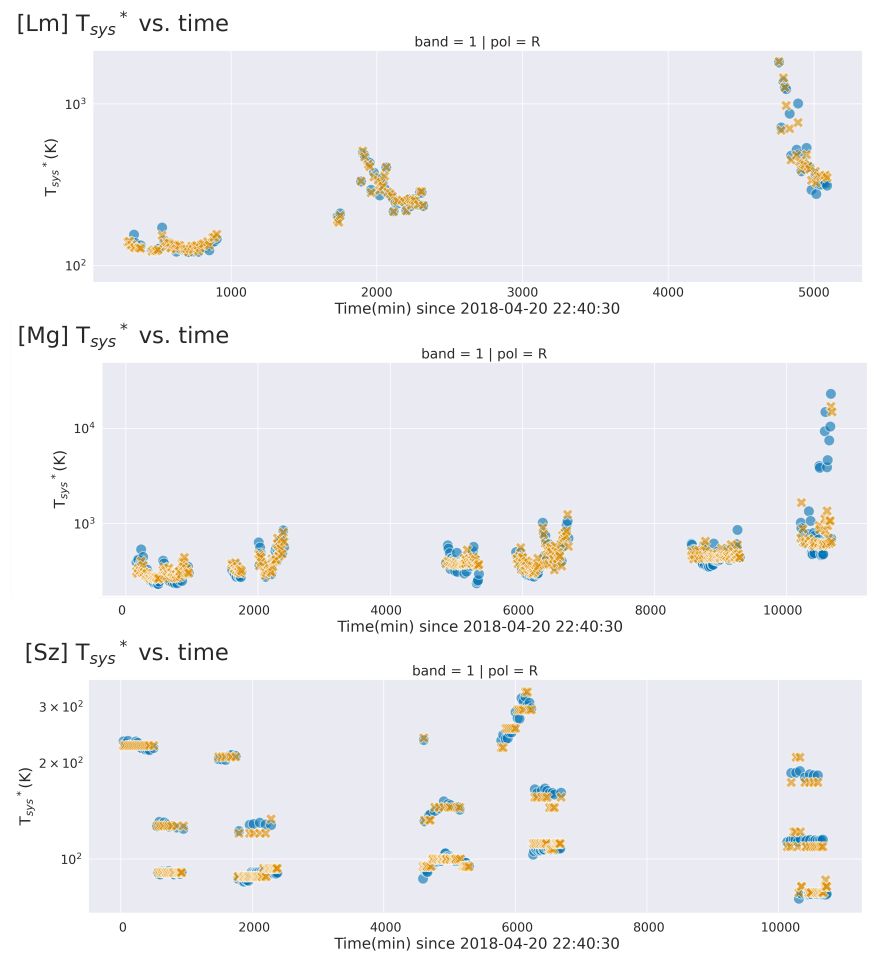}
\caption{Comparison between original $T_{\rm sys}^{*}$ values (blue circles) and modelled ones (orange crosses) for antennas LMT (top), SMT (middle), and SPT (bottom) in RCP, band 1. The latter two antennas are the ones for which the model is less accurate. This could be related to significant time variations in $T_{\rm atm}$ and $T_{\rm amb}$ within a track, or due to directional or time dependent changes in opacity. However, as a first approximation and with the aim of finding missing $T_{\rm sys}^{*}$ values, we consider this approach as reasonable. Note that original values with flag code ``U'' do not have a modelled counterpart (this is more noticeable in the last observing track for SMT). }\label{fig:fit}
\end{figure*}

\begin{figure*}[!ht]
\centering
\includegraphics[width=0.95\textwidth]{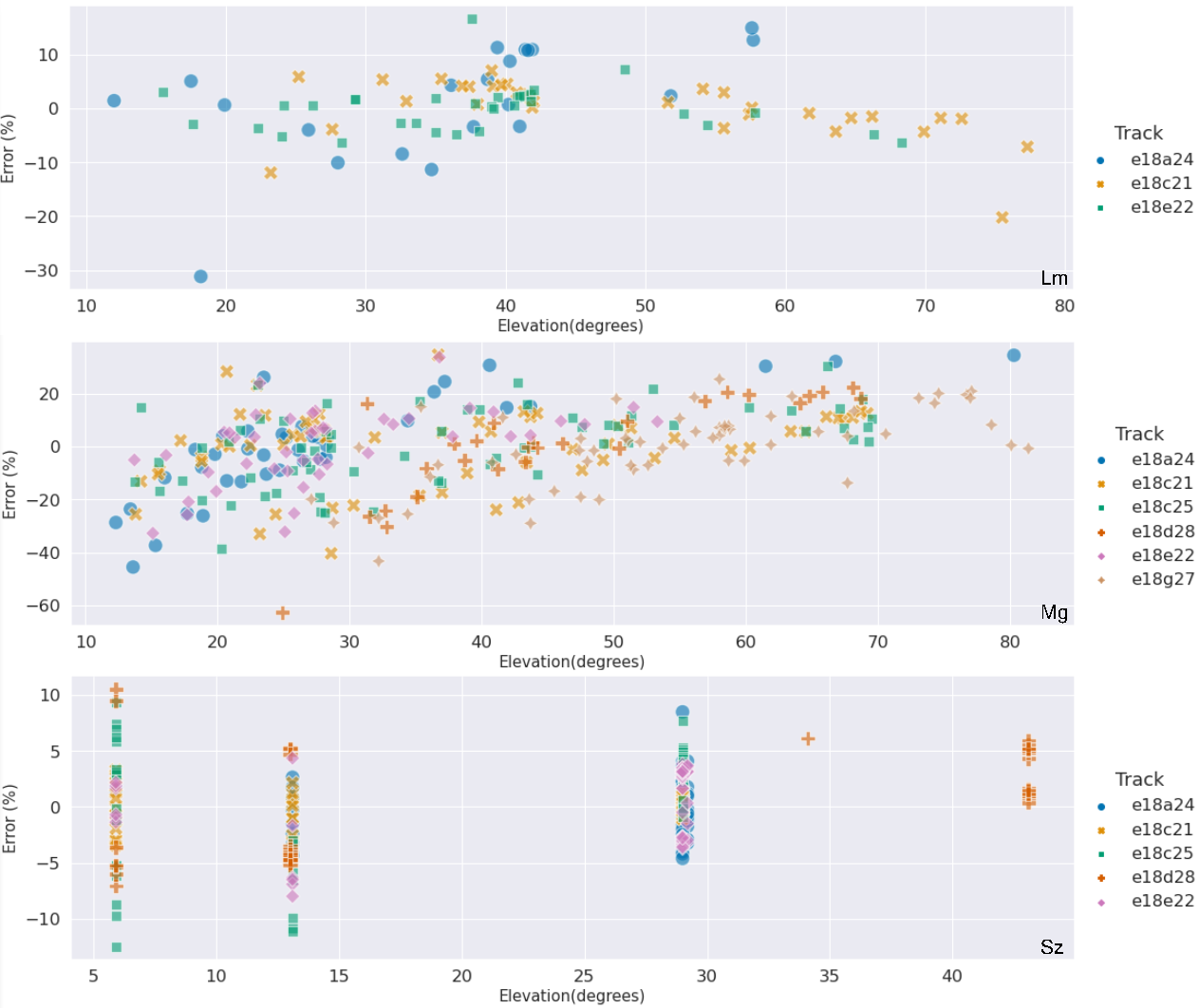}
\caption{Error percentage between $T_{\rm sys}^{*}$ modelled and original values vs elevation, for antennas LMT (top), SMT (center), and SPT (bottom) in RCP, band 1. SMT in particular had abrupt changes in  $T_{\rm sys}^{*}$ which cannot be handled properly by the fit, therefore resulting in larger error percentages compared to other antennas. The largest error ($>60$\,\%) occurs in track e18d28, scan No0033, which comes right after a scan with flag code ``U'' on the same source.}\label{fig:fit_error}
\end{figure*}

Missing values in the original single-dish tables (e.g., VEX scans that were observed but are missing $T_{\rm sys}^{*}$ counterparts, and scans deemed unreliable and thus flagged) are modelled by curve-fitting equation \ref{eq:tsysmod} to the available $T_{\rm sys}^{*}$ values in each track. Once $Q_0$ and $Q_1$ are obtained, we applied them to the scans missing $T_{\rm sys}^{*}$, using their corresponding elevation and opacity values from the metadata tables, or when missing, using the calculated elevation based on observing time, source position and antenna location, and the interpolated opacity from neighbouring scans. We note that the fitted parameters $Q_0$ and $Q_1$ in equation \ref{eq:tsysmod}, are obtained per track and per antenna while excluding those values with flag code ``U''. This approximation does not allow for an exact representation of $T_{\rm sys}^{*}$ in time, given that $Q_0$ and $Q_1$ (which incorporate quantities such as $T_{\rm atm}$, $T_{\rm amb}$ and $\eta_{\rm l}$) are treated as constants for each combination of track and antenna, although strictly speaking, they can vary in time. This representation results in modeled $T_{\rm sys}^{*}$ values which are close enough to the original ones for most of the metadata (see Figure \ref{fig:fit}). However, we note that the model performs poorly for scans in tracks where the $T_{\rm sys}^{*}$ changes rapidly for a given source (see Figure \ref{fig:fit_error}).

\subsection{Additional Processing for Phased Arrays}\label{sec:process_phased}

For the phased arrays, i.e. ALMA and SMA, the DPFUs and phasing efficiencies are factored into the $T_{\rm sys}^{*}$ values in the ANTAB calibration tables. Therefore, for these two stations, the $T_{\rm sys}^{*}$ sections of the ANTAB tables in fact contain their SEFD values as a function of time. 

\subsubsection{ALMA SEFDs}\label{sec:almasefd}

For ALMA, ANTAB format calibration tables are provided by the team responsible for the QA2 and {\fontfamily{qcr}\selectfont Polconvert} processing of the ALMA EHT data. For each track and observing band, the tables contain 32 columns of SEFD values (with 0.031\,K/Jy factored out as a single-dish DPFU), one for each of the 32 spectral windows. 

These SEFDs are provided at a rate of less than a second (with rates as high as 0.4s), interpolated by {\fontfamily{qcr}\selectfont Polconvert} to match the time resolution of the VLBI correlation times. 

However, since the {\fontfamily{qcr}\selectfont AIPS} software is unable to handle such a high cadence, our processing pipeline averages the values over 6s before writing them into the combined ANTAB table for all stations.

\subsubsection{SMA SEFDs}\label{sec:smasefd}

The SMA provides DSB $T_{\rm sys}^{*}$ measurements for each antenna and each polarization, measured at a cadence of once every minute. The phasing efficiencies are provided for each band and polarization, at a shorter cadence of 20 to 30s per measurement. These data are processed via the following steps:
\begin{enumerate}
\item The SMA $T_{\rm sys}^{*}$ values are multiplied by a factor 2 to obtain the SSB equivalent values \citep[see][]{issaounetal17a}, for a sideband ratio of 1. 

\item The $T_{\rm sys}^{*}$ values are then divided by the DPFUs of the corresponding track, antenna and polarization, to convert them from units of K to units of Jy. Since the DPFU values for each band are roughly comparable, and the $T_{\rm sys}^{*}$ values are the same for all 4 bands, we average the DPFU values across all 4 bands as well. 

\item The phased $T_{\rm sys}^{*}$ (in units of Jy) over all 7 antennas is then estimated following Equation~3 in the memo by \cite{janssenetal19}.
\item The SEFDs for each band and polarization are estimated by dividing the phased $T_{\rm sys}^{*}$ by the corresponding phasing efficiencies. The phasing efficiencies are first averaged over the one minute cadence of the $T_{\rm sys}^{*}$ measurements.
\item SEFD outlier values and those measured during gaps in observing scans (e.g., when the antennas are slewing) are flagged.
\end{enumerate}

\subsection{Validation}\label{ssec:val_plots}

\begin{figure*}[ht]
\centering
\includegraphics[width=0.99\textwidth]{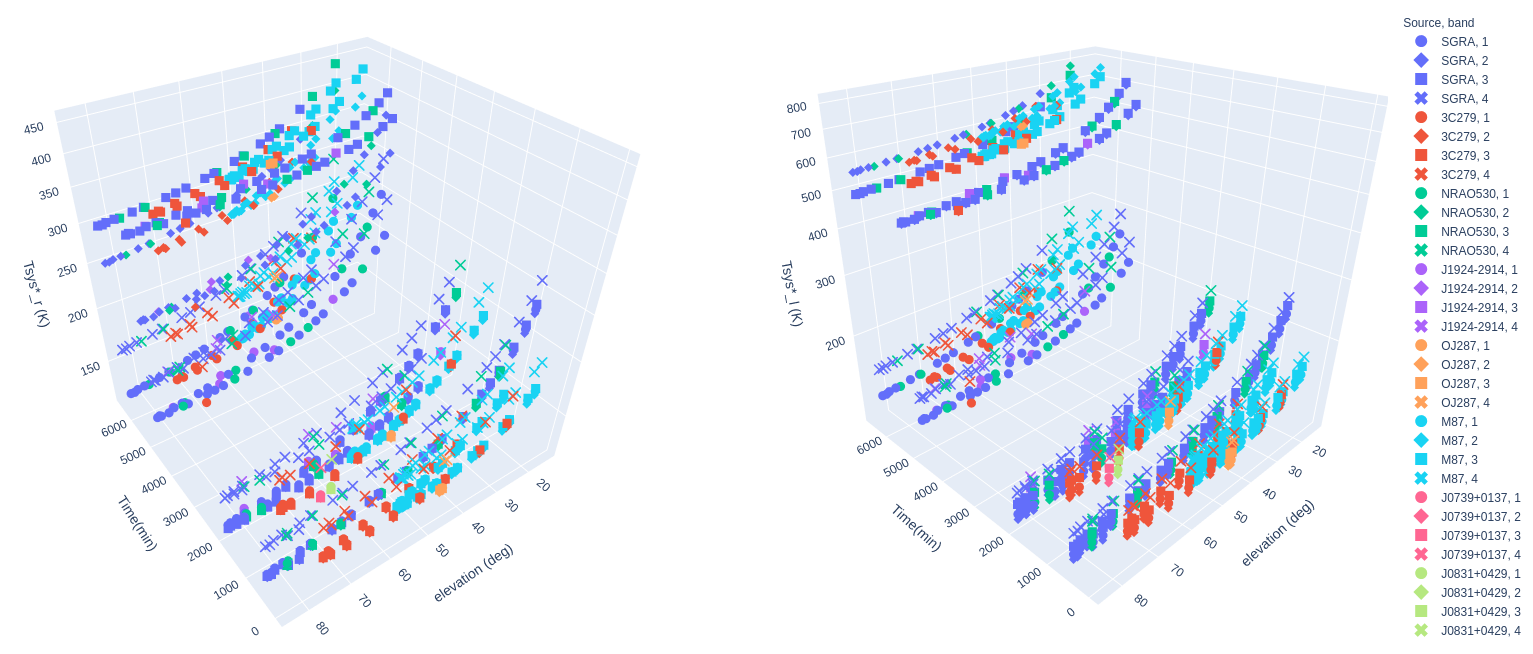}
\caption{Snapshots of interactive 3D plots for APEX $T_{\rm sys}^{*}$ values as a function of both time and elevation for RCP (left panel) and LCP (right panel). Each source is marked with a different color, and each frequency band is denoted by a different marker shape.}\label{fig:3d-apex}
\end{figure*}

\begin{figure*}[ht]
\centering
\includegraphics[width=0.99\textwidth]{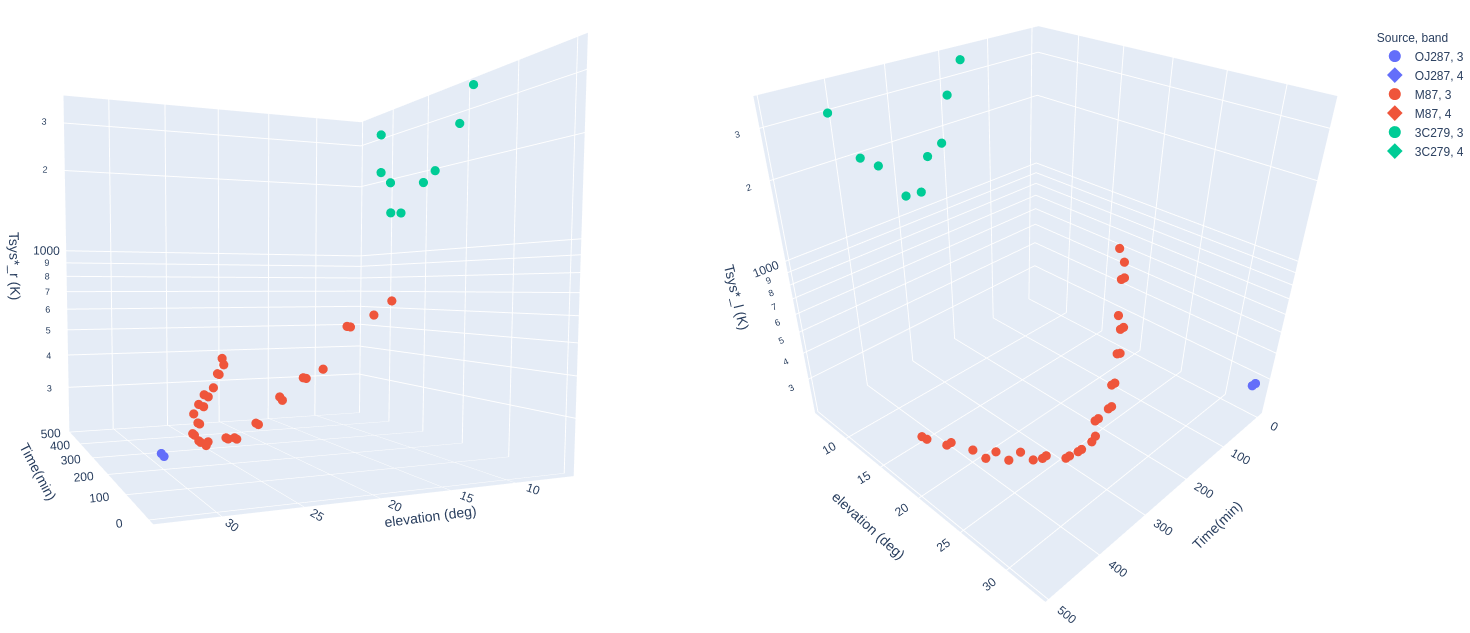}
\caption{Snapshots of interactive 3D plots for GLT similar to those in Figure \ref{fig:3d-apex}. Here we isolate track e18c21 to visualize clearly how $T_{\rm sys}^{*}$ in both RCP and LCP, vary with time and elevation. Band 3 is represented by circles, and band 4 by diamonds. The observed sources in the isolated track are OJ287 (blue), M87 (red) and 3C279 (green).}\label{fig:3d-glt_c21}
\end{figure*}

The pipeline (Section \ref{sec:pipeline}) generates 2D and 3D plots (using the  {\fontfamily{qcr}\selectfont Seaborn} and  {\fontfamily{qcr}\selectfont Plotly} data visualization libraries in {\fontfamily{qcr}\selectfont python}) which allow us to visually assess the quality of the data from single-dish antennas. The validation plots are provided per antenna, colour-coded by target source and symbol-coded by polarization or observing track: (1) 2D plots: opacity vs. time and $T_{\rm sys}^{*}$ vs. time. (2) 3D interactive plots of $T_{\rm sys}^{*}$ vs. elevation and time, per polarization. The 3D interactive plots are particularly useful because they allow a direct identification of the origin of suspicious $T_{\rm sys}^{*}$ values by hovering the mouse above them, after which annotations become visible. By double-clicking on particular labels, one can also isolate different tracers, thus enhancing the visibility of different data at a time. Here we show some examples based on APEX metadata for all the 2018 tracks (Figure \ref{fig:3d-apex}), and GLT metadata for track e18c21 only (Figure \ref{fig:3d-glt_c21}). All the validation plots can be found in the L1 metadata tarball (see Section~\ref{sec:l1products} for details). 

Similar 2D validation plots of $T_{\rm sys}^{*}$, phasing efficiencies and SEFDs as a function of time and elevation are also produced for the phased arrays, and are included in the metadata package.

\subsection{ANTAB Table Generation}\label{sec:antabgen}

Having processed the metadata as described above, and having calculated the DPFU values and gain curve coefficients, we created ANTAB tables in the standard format\footnote{\url{http://www.aips.nrao.edu/cgi-bin/ZXHLP2.PL?ANTAB}} \citep[see also][for detailed descriptions]{janssenetal19}. The DPFU values and gain curve polynomial coefficients are attached as headers, and the $T_{\rm sys}^{*}$ for single dish stations and SEFD values for phased arrays, ordered by station, are attached in the main body. This process generates one ANTAB table per observing track, per band. We produce two sets of ANTAB tables: 
\begin{enumerate}
\item \textbf{raw ANTAB tables}: contains single-dish station $T_{\rm sys}^{*}$ values without any flagging or corrections applied, using the original timestamps provided by the stations. The only `processing' steps applied to this version are for the determination of the phased array SEFDs (as described in Section~\ref{sec:process_phased}), the conversion of JCMT $T_{\rm sys}^{*}$ values from DSB to SSB equivalents, and the factoring of the JCMT time-dependent gains into the $T_{\rm sys}^{*}$ values.
\item \textbf{processed ANTAB tables}: In addition to the processing of the phased-array SEFDs and JCMT $T_{\rm sys}^{*}$ as done for the raw ANTAB tables, this version contains $T_{\rm sys}^{*}$ values for which different flagging steps have been applied, and for which observed scans with missing $T_{\rm sys}^{*}$ values have been replaced with modelled ones. In addition, a shift in the timestamps has been applied (to coincide with the center of each scan) for single dish stations. 
\end{enumerate}

We expect project PIs and the EHT L2 calibration teams to mainly use the processed ANTAB tables for calibration. The raw ANTAB tables can be used for diagnostic checks.

\section{L1 Metadata Package: Contents and Data Properties}\label{sec:l1products}

\subsection{Package Contents}\label{sec:contents}

As part of the 2018 L1 data release, the metadata is packaged as a single tarball (version {\fontfamily{qcr}\selectfont X.X}): \\ 
{\fontfamily{qcr}\selectfont EHTmetadata\_2018April\_vX.X}\\

This tarball package contains the following products each in a separate directory: 

\begin{itemize}
    \item \textbf{ANTAB calibration tables}, separated by tracks and frequency bands. Each table contains DPFUs and gain curve coefficients for all stations in the header, and $T_{\rm sys}^{*}$ values as a function of time for all single-dish stations, and SEFD values as a function of time for both ALMA and the SMA. As mentioned above, two versions of ANTAB tables are provided for each track and band, (1) the raw ANTAB files in the directory {\fontfamily{qcr}\selectfont antab/raw/}; and (2) the processed ANTAB files in the directory {\fontfamily{qcr}\selectfont antab/processed/}.

    \item \textbf{Validation plots} used for quality assessment, as described in Section~\ref{ssec:val_plots}. They are located in the {\fontfamily{qcr}\selectfont plots/} directory.
    
    \item \textbf{Auxiliary files} produced by the metadata processing pipeline (Section~\ref{sec:pipeline}), including tables of flagged or problematic $T_{\rm sys}^{*}$, tables comparing raw and modeled $T_{\rm sys}^{*}$, as well as {\fontfamily{qcr}\selectfont python} dictionaries of names and codes for tracks, stations and sources. Descriptions of these individual files are provided in the README contained within the directory {\fontfamily{qcr}\selectfont aux/}, where these files are also located.
    
    \item \textbf{Raw metadata files} collected from each station, as described in Section~\ref{sec:rawcollect}. They are located in the directory {\fontfamily{qcr}\selectfont raw/}.
    
\end{itemize}
 
This metadata package is available online in the public domain \footnote{\url{https://doi.org/10.25739/2x5r-xd34}}. The tarball is also staged and archived together with the correlator products for the 2018 observing campaign. For reference, tables summarizing identification codes used in the naming of files associated with the various observing stations and tracks are provided in Appendix~\ref{app:ids}.

\subsection{Comparing 2017 and 2018 Station Sensitivities} \label{sec:comparisons2017}

The sensitivity parameters of each station in 2018 are summarised in Table~\ref{tab:sensitivities}, in comparison with their values in 2017. A few notable differences include:
\begin{itemize}
    \item The median $T_{\rm sys}^{*}$ values are higher at ALMA, APEX, SMT and IRAM 30\,m stations, due to overall poorer weather in 2018 compared to 2017.
    \item The LMT sensitivity has improved in the 2018 observations, after the upgrade from a 32.5\,m to a 50\,m diameter dish. We note also that the LMT switched to 2SB receivers just before the April 2018 EHT campaign; the telescope was operating with an interim 1mm DSB receiver in the 2017 which has since been decommissioned. 
    \item The IRAM 30\,m telescope DPFU has also increased in 2018, after the installation of a new maser. Amplitude losses due to decoherence, attributed to excess noise in the maser frequency reference in 2017, required an ad-hoc down-scaling of the telescope DPFU by a factor of 3.6.
\end{itemize}

\begin{table*}
\caption{Median sensitivities for primary targets in the EHT 2017 \citep[from][]{ehtpaperIII19} and 2018 campaigns. The antenna efficiencies ($\eta_{\rm A}$) and DPFUs are aggregated over frequency bands and polarizations, where necessary.}\label{tab:sensitivities}
\begin{tabularx}{\linewidth}{@{\extracolsep{\fill}} l c c c c c c c c c c}
\hline\hline       
Station & \multicolumn{2}{c}{diameter (m)} &  \multicolumn{2}{c}{Median $T_{\rm sys}^{*}$ (K)} & \multicolumn{2}{c}{$\eta_{\rm A}$ (\%)} & \multicolumn{2}{c}{DPFU (K/Jy)} & \multicolumn{2}{c}{Median SEFD (Jy)} \\ 
\cline{2-3}\cline{4-5}\cline{6-7}\cline{8-9} \cline{10-11}
 & 2017 & 2018 & 2017 & 2018 & 2017 & 2018 & 2017 & 2018 & 2017 & 2018 \\ 
\hline  
ALMA & 73\textsuperscript{a} (37) & 79\textsuperscript{a} (43) &  76 &  113 & 68\textsuperscript{b} & 68\textsuperscript{b} & 1.030 & 1.190  &    74  &   106 \\    
APEX & 12  & 12  &  118 &  163 & 61 & 64 &  0.025  & 0.026  &  4,800  &  6,259 \\
GLT & - & 12     &   -  &  410 & - & 22 &  -           & 0.0088       &   -    & 46,287 \\
LMT & 32.5 & 50  &  371 &  208 & 28 & 22 &  0.083       & 0.157       &  4,500  &  1,362 \\
SMT & 10 & 10    &  291 &  375 & 60 & 65 &  0.017       & 0.019       & 17,100  & 20,330 \\
JCMT & 15 & 15    &  345\textsuperscript{c} &  379\textsuperscript{c} & 52 & 46 &  0.033       & 0.030       & 10,500  & 12,816 \\
IRAM 30\,m & 30 & 30 & 226 &  351 & 47 & 55 & 0.034\textsuperscript{d}       & 0.140       &  6,900  &  2,539 \\
SMA & 14.7\textsuperscript{a} (6) & 15.9\textsuperscript{a} (7) &  285\textsuperscript{c} & 283\textsuperscript{c} & 75\textsuperscript{b} & 75\textsuperscript{b} &  0.046   & 0.040  &  6,400  &  8,458 \\
SPT & 10 & 10 &  118 &   95 & 60\textsuperscript{e} & 25 & 0.0061 & 0.0071  & 19,300  & 13,365 \\
\hline
\end{tabularx}
\newline
\textsuperscript{a}\footnotesize{The diameter for phased arrays reflects the total collecting array of all antennas. Values in parentheses represent the number of participating antennas.}\newline
\textsuperscript{b}\footnotesize{The quoted aperture efficiency for phased arrays are that for each single dish.}\newline
\textsuperscript{c}\footnotesize{The quoted $T_{\rm sys}^{*}$ are SSB equivalent values, assuming sideband ratios of 0.9 for JCMT and 1.0 for SMA.}\newline
\textsuperscript{d}\footnotesize{A correction factor of 3.663 was applied to the 2017 IRAM 30\,m DPFU, due to losses from maser frequency instabilities.}\newline
\textsuperscript{e}\footnotesize{The SPT antenna efficiency for 2017 was estimated assuming a 6m diameter dish, considering only 6m illumination of the dish by the receiver optics. While the antenna is still under-illuminated in 2018, the antenna efficiencies in 2018 are estimated with respect to the full 10\,m dish.}\newline
\end{table*}

\subsection{Remaining Issues in the {\fontfamily{qcr}\selectfont EHTmetadata\_2018April\_v1.0} Package} \label{sec:dataissues}

We summarize here outstanding issues in the release of the {\fontfamily{qcr}\selectfont EHTmetadata\_2018April\_v1.0} package, that may affect the a-priori amplitude calibration of EHT 2018 data. More careful analysis is required downstream to mitigate these issues. If possible and required, they will be corrected in an updated version of the metadata package. 
\begin{itemize}
    \item There remains a large scatter in the SMA SEFDs on some tracks (e.g., at the beginning of tracks e18c25 on April 25, e18g27 on April 27, and almost the entire track of e18d28 on April 28). Extreme SEFD outliers have been flagged, but there are borderline cases where more careful flagging and analyses are needed after applying the a-priori calibration to fringe-fitted data. Ad-hoc SEFD corrections may also be needed on some range of scans, as was done for 2017 SMA data.
    \item Due to very poor weather (including periods of snow) at the ALMA site on tracks e18g27 (April 27) and e18d28 (April 28), the QA2 results are less reliable. For example, as a result of poor phasing, there are some jumps in the ALMA SEFDs on some scans, which may need to be flagged pending further analysis. 
    \item As mentioned, the LMT gain curve is not yet well-characterized. Large gain errors of up to 20\% (or possibly higher) are expected at low elevation scans, which may need to be corrected downstream via self-calibration or other methods. The gains can also be determined in the future once the telescope operations resume.
    \item The APEX DPFUs may need an ad-hoc rescaling (or network calibration) to account for an underestimation of flux densities on APEX baselines, when compared to the flux densities of similar ALMA baselines to the same stations. Additionally, for band 1, we found 25\% differences between RCP and LCP amplitudes on all APEX baselines. For band 4, there is a smaller but still significant 13\% difference between the RCP and LCP amplitudes. These issues were discovered after conducting preliminary a-priori calibration tests. Follow-up diagnostics found this issue to be caused by low IF power levels and too large block-down converter attenuation settings for the R2DBE associated with RCP in bands 1 and 4.
    \item Due to the small number of measurements, the GLT DPFU uncertainties are not well characterized, and are likely to be larger than the quoted 7\%. $T_{\rm sys}^{*}$ uncertainties are also large, $\sim$15\%. Pointing errors at low elevations in 2018 can cause amplitude losses of up to 18\% at $10^{\circ}$ elevation, up to as much as 96\% at $6^{\circ}$ \citep{koayetal23}; this affects scans observed at elevations below $10^{\circ}$, e.g. for 3C279.
    \item The frequency dependence of $T_{\rm sys}^{*}$ at the JCMT is unknown, and we have simply applied the $T_{\rm sys}^{*}$ measured in bands 2 and 3 to bands 1 and 4. Network calibration may be able to correct for these errors in bands 1 and 4.
\end{itemize}

\subsubsection{Minor updates in the {\fontfamily{qcr}\selectfont EHTmetadata\_2018April\_v1.1} Package}

We have made minor modifications in v1.1 of the metadata package.  These minor modifications include:
\begin{itemize}
    \item Adding the raw JCMT (Mm) metadata files, which were missing in the v1.0 of the release. The new raw metadata release now includes all the files provided by the telescope representatives.
    \item Using updated sideband ratios measured in 2018 to convert JCMT DSB $T_{\rm sys}^{*}$ values into their SSB ones equivalents (as described in section \ref{sec:sd_individual}). Previously we used the sideband ratio of from 2017 \citep[see][]{issaounetal17a}.
    \item Updating the validation plots and SEFD tables provided in the auxiliary files. These were affected by a bug in the processing pipeline which did not implement properly the needed updates after flagging and removing extreme outliers in the SMA (Sw). This bug did not affect any of the ANTAB tables used for calibration.
\end{itemize}

All the issues described above for v1.0 are still present in v1.1.

\newpage

\appendix

\section{Naming Conventions and Identifications for Stations and Tracks}\label{app:ids}

We provide a summary of the naming conventions for telescopes (Table~\ref{tab:stations}) and observing tracks (Table~\ref{tab:tracks}) used in this memo and in the naming of files in the metadata package. We show also the single-letter HOPS station code, used in the HOPS data reduction pipeline as well as many custom EHT data validation and plotting libraries.

\begin{table}[h]
\caption{Names and identifications for all the telescopes that participated in the 2018 EHT observing campaign.}\label{tab:stations}
\begin{tabularx}{\linewidth}{@{\extracolsep{\fill}}l c c c c}
\hline\hline       
Full Telescope Name & Abbreviation & VEX Site Name & VEX Code & HOPS Code\\
\hline 
Atacama Large Millimeter/submillimeter Array &  ALMA & ALMA & Aa & A\\    
Atacama Pathfinder Experiment  & APEX & APEX & Ax & X \\
Greenland Telescope & GLT & THULE & Gl & G \\
Large Millimeter Telescope & LMT & LMT & Lm & L\\
Submillimeter Telescope & SMT & SMTO & Mg & Z\\
James Clerk Maxwell Telescope & JCMT & JCMT & Mm & J\\
Institut de Radioastronomie Millim\'{e}trique 30\,m Telescope & IRAM 30\,m & PICOVEL & Pv & P \\
Submillimeter Array & SMA & SMAP & Sw & S\\
South Pole Telescope & SPT & SPT & Sz & Y\\
\hline
\end{tabularx}
\end{table}

\vskip 1cm

\begin{table}[h]
\caption{VEX and EHT experiment identification numbers for each track of the 2018 EHT observing campaign. Participating stations for each track are also shown.}\label{tab:tracks}
\begin{tabularx}{\linewidth}{@{\extracolsep{\fill}}l c c l}
\hline\hline       
Observing Date\textsuperscript{a} (UT) & VEX ID & EHT Experiment No. & Participating Stations  \\
\hline 
2018 Apr 21 & e18c21 & 3644 & Aa, Ax, Lm, Pv, Sw, Mm, Mg, Gl, Sz \\
2018 Apr 22 & e18e22 & 3645 & Aa, Ax, Lm,     Sw, Mm, Mg, Gl, Sz\\
2018 Apr 24 & e18a24 & 3646 & Aa, Ax, Lm, Pv, Sw, Mm, Mg,          Sz\\
2018 Apr 25 & e18c25 & 3647 & Aa, Ax,     Pv, Sw, Mm, Mg, Gl, Sz\\
2018 Apr 27 & e18g27 & 3648 & Aa,         Pv, Sw, Mm, Mg, Gl\\
2018 Apr 28 & e18d28 & 3649 & Aa,         Pv, Sw, Mm, Mg, Gl, Sz\\

\hline
\end{tabularx}
\newline
\textsuperscript{a}\footnotesize{These dates correspond to the day on which the majority of the track runs. Some of the tracks actually start a few hours prior to the start of specified date in UT.}\newline

\end{table}

\begin{landscape}
\section{Formats and Contents of Raw Metadata Tables}\label{app:rawformat}

\quad Example $T_{\rm sys}^{*}$ table format and contents.\\ 
\quad Filename: \texttt{e18c21\_SZ.tsys}
\begin{Verbatim}[fontsize=\footnotesize]
################################################################
#
# Tsys/Tsys* table for each station and track
#
################################################################
#
# Track Start Date (UT): 2018-04-20
# Experiment name: e18c21
# Station ID: SZ
# Operators/observer/contact person: eventhorizon telescope (eht@eht.edu)
# Tsys* or Tsys (inclusive of opacity or not): Tsys*
# Central observing frequencies in GHz (DSB/2SB): 213.1, 215.1, 227.1, 229.1, (2SB)
# Sideband ratio (if available, for DSB): NA
# Sideband coupling coefficient (if available, for 2SB): 
# Tau observing frequency in GHz: 225.0
#
################################################################
# * Timestamp gives the date and time at which Tsys*/Tsys was measured (can be during, before or after each scan)
# * Scan column gives the VEX scan number associated with each Tsys*/Tsys measurement 
# * Tau values should be that measured at zenith
# * Empty cells should have an 'NA'
# * If the same Tsys/Tsys* values are used for multiple bands, then include repeated values in the appropriate columns
# * Tsys*/Tsys values should be provided without sideband correction
#
# Timestamp(UT)	  Scan   Source   Pos_Az Pos_El  Tsys_b1r Tsys_b1l Tsys_b2r Tsys_b2l Tsys_b3r Tsys_b3l Tsys_b4r Tsys_b4l    Tau      Tamb   Tatm
# YYYY-MM-DD HH:MM:SS   (VEX)	    (deg)  (deg)      (K)     (K)      (K)      (K)      (K)      (K)      (K)      (K)     (zenith)   (K)    (K)
##########################################################################################################################################################
2018-04-21 06:51:21     No0051  3C279    240.2    5.9    222.6    218.5    226.3    223.0    218.5    214.8    216.6    209.2     0.053     271.7  220.9
2018-04-21 07:49:57     No0054  NRAO530  296.4   13.1    126.6    124.1    129.5    129.2    126.2    123.8    127.3    122.9     0.054     254.0  221.0
2018-04-21 07:58:18     No0055  SGRA     297.4   29.0     90.2     88.6     92.6     93.8     91.1     88.8     93.8     89.8     0.054     261.3  220.8
2018-04-21 08:10:04     No0056  SGRA     294.1   29.0     89.6     87.8     91.8     93.3     90.5     88.3     92.9     89.3     0.054     265.0  220.7
2018-04-21 08:22:41     No0057  NRAO530  287.7   13.1    130.8    128.1    133.9    134.2    130.3    128.0    131.2    127.0     0.052     268.0  220.8
\end{Verbatim}

\newpage
\quad Example flag table format and contents.\\ 
\quad Filename: \texttt{e18c21\_SZ.flag}
\begin{Verbatim}[fontsize=\footnotesize]

################################################################
# 
# Flag tables
#
################################################################
#
# Track Start Date (UT): 2018-04-20
# Experiment name: e18c21 
# Station ID: SZ
# Operators/observer/contact person: eventhorizon telescope (eht@eht.edu)
# Known bad IFs and channels:
#
################################################################
# * Scan column is the VEX scan number 
# * ScanStart and ScanStop are start and stop times for each scan following VEX file
# * FlagCode - assign appropriate code(s) for each scan, using multiple if necessary (e.g. PT, UT):
#   	S: Success, no problems
#   	N: Not observed and data totally unusable (total flag of scan, use this only if 100% sure)
#   	P: Partial observation (flag partial time-range) e.g late on-source
#   	U: Not so good or uncertain data quality (keep for now, note for downstream users) e.g., bad pointing
#   	T: Missing Tsys*/Tsys for this scan (interpolation from other scans required)
# * Limit operator comments to 100 characters
#
################################################################ 
# 
#
# Scan   ScanStart(UT)	      ScanStop(UT)		Source	FlagCode   Observer/Operator Comments
# (VEX)  YYYY-MM-DD HH:MM:SS  YYYY-MM-DD HH:MM:SS
##################################################################################################
No0051   2018-04-21 06:59:00  2018-04-21 07:03:00       3C279      S        #
No0052   2018-04-21 07:23:00  2018-04-21 07:33:00       SGRA       T        # Tsys scan at wrong elevation.
No0053   2018-04-21 07:36:00  2018-04-21 07:46:00       SGRA       TP       # No Tsys scan. Late on scan.
No0054   2018-04-21 07:51:00  2018-04-21 07:55:00       NRAO530    S        #
No0055   2018-04-21 08:00:00  2018-04-21 08:10:00       SGRA       S        #
No0056   2018-04-21 08:13:00  2018-04-21 08:22:00       SGRA       S        #
No0057   2018-04-21 08:26:00  2018-04-21 08:30:00       NRAO530    S        #

\end{Verbatim}

\newpage
\quad Example phasing efficiency table format and contents.\\ 
\quad Filename: \texttt{e18c21\_SW.pheff}
\begin{Verbatim}[fontsize=\footnotesize]
################################################################
#
# Phasing efficiencies for each phased array and track
#
################################################################
# Track Start Date (UT): 2018-04-21
# Experiment name: e18c21
# Station ID: SW
# Operators/observer/contact person: eventhorizon telescope (eht@eht.edu)
#################################################################
# *Timestamp gives time for each phasing efficiency measurement
# *Cadence and precision of phasing efficiencies should follow that produced by the station
# *N is the number of telescopes observing in the array
#
# Timestamp(UT)        N  peff x 4 bands x 2 pols
# YYYY-MM-DD HH:MM:SS     Band-1-RCP  Band-1-LCP  Band-2-RCP  Band-2-LCP  Band-3-RCP  Band-3-LCP  Band-4-RCP  Band-4-LCP
##############################################################################################################################
2018-04-21 05:56:09    7  0.24024871  0.05376614  0.26618701  0.10487495  0.18657796  0.37746889  0.24862157  0.18657796
2018-04-21 05:56:14    7  0.07714367  0.14246480  0.32104154  0.13491297  0.07742733  0.37199276  0.23416090  0.07742733
2018-04-21 05:57:02    7  0.31014805  0.05887395  0.15769393  0.12820571  0.05147650  0.41022153  0.27828420  0.05147650
2018-04-21 05:57:51    7  0.50892541  0.17867515  0.07574473  0.30626215  0.13132703  0.56593394  0.53427692  0.13132703
2018-04-21 05:58:21    7  0.38716008  0.12616070  0.16030566  0.20038906  0.10372790  0.64648049  0.59182322  0.10372790
2018-04-21 05:58:51    7  0.15929065  0.18521967  0.26011389  0.24312429  0.03929938  0.84489059  0.79754373  0.03929938
2018-04-21 05:59:20    7  0.23347205  0.13258306  0.26167665  0.21430481  0.11962959  0.54098511  0.40466134  0.11962959
2018-04-21 05:59:50    7  0.14592671  0.13400968  0.35303846  0.20403350  0.18310070  0.47837284  0.35911362  0.18310070
2018-04-21 06:00:20    7  0.29739306  0.12892525  0.21754603  0.21462931  0.14821269  0.66865731  0.60594267  0.14821269
2018-04-21 06:00:49    7  0.20361152  0.19952152  0.19580552  0.25275614  0.06050972  0.83088507  0.82691807  0.06050972
2018-04-21 06:01:19    7  0.11900591  0.10304374  0.33619428  0.14566428  0.11043349  0.55654936  0.41239379  0.11043349
2018-04-21 06:01:49    7  0.25732087  0.76051137  0.84108035  0.79557787  0.72244056  0.86797338  0.79854751  0.72244056
2018-04-21 06:02:18    7  0.72950628  0.71199815  0.62463551  0.75040914  0.69852207  0.71572688  0.66894971  0.69852207
2018-04-21 06:02:48    7  0.62666174  0.37809111  0.57887363  0.62199385  0.67937367  0.62098385  0.60202166  0.67937367
\end{Verbatim}

\newpage
\quad Example gains table format and contents.\\ 
\quad Filename: \texttt{SZ\_2018-04.gc}
\begin{Verbatim}[fontsize=\footnotesize]

################################################################
# Gain curve table and measurements
################################################################
#
# Range of dates of measurements: 2018-03-20 to 2018-10-05
# Station ID: SZ
# Antenna Mount Type: Offset Gregorian, Alt-Az mount
# Antenna Diameter (m): 12
# Operators/observer/contact person: eventhorizon telescope (eht@eht.edu)
#
#################################################################
# Timestamp gives the date and time for each antenna efficiency measurement
# Freq is the observing frequency
# errAeff_r and errAeff_l are the errors/uncertainties of the measured antenna efficiencies
# Tau should be the value measured at zenith
#
# Aeff_r and Aeff_l from if0 and if1 channels, respectively.
# Timestamp(UT)       Freq   Source    El  Aeff_r  errAeff_r  Aeff_l  errAeff_l   Tsys*   Tau   Tamb   Tatm   Notes/Comments
# YYYY-MM-DD HH:MM:SS (GHz)          (deg)                                        (K)  (zenith)  (K)   (K)
############################################################################################################################
2018-04-28  13:03:15   221   Saturn   22.2  0.253    0.03      0.263     0.03     83.8   0.046  247.7  207.6
2018-04-28  13:38:43   221   Saturn   22.2  0.250    0.03      0.263     0.03     83.8   0.046  246.7  207.6
2018-04-28  15:00:00   221   Jupiter  16.4  0.236    0.02      0.247     0.02     94.0   0.047  248.7  207.7
2018-04-28  15:33:26   221   Jupiter  16.4  0.242    0.02      0.255     0.02     94.0   0.047  247.7  207.5

\end{Verbatim}

\end{landscape}


\begin{thebibliography}{}
\bibitem[Butler(2012)]{butler12} Butler, B. 2012, ALMA Memo 594
\bibitem[Event Horizon Telescope Collaboration et al.(2019)]{ehtpaperIII19} Event Horizon Telescope Collaboration, et al., 2019, ApJL 875, L3
\bibitem[Goddi et al.(2019)]{goddietal19} Goddi, C., Mart\'{\i}-Vidal I., Messias, H., et al. 2019, \pasp, 131, 075003 
\bibitem[Greisen(2003)]{greisen03} Greisen, E. W. 2003, in Information Handling in Astronomy-Historical Vistas, Vol. 285, ed. A. Heck (Dordrecht: Kluwer), 109
\bibitem[Inoue et al.(2014)]{inoueetal14} Inoue, M., Algaba-Marcos, J. C., Asada, K., et al. 2006, Radio Science, 49, 564	
\bibitem[Chen et al.(2023)]{chenetal23} Chen M.-T., Asada K., Matsushita S., Raffin P., Inoue M., Ho P.~T.~P., Han C.-C., et al., 2023, PASP, 135, 095001
\bibitem[Issaoun et al.(2017a)]{issaounetal17a} Issaoun S., Folkers, T. W., Blackburn, et al. 2017, EHT memo series, 2017-CE-02 \\ \url{https://eventhorizontelescope.org/files/eht/files/EHT_memo_Issaoun_2017-CE-02.pdf}
\bibitem[Issaoun et al.(2017b)]{issaounetal17b} Issaoun S., Folkers, T. W., Marrone, D. P., Kim, J., Tilanus, R., and Falcke, H. 2017, EHT memo series, 2017-CE-03 \url{https://eventhorizontelescope.org/files/eht/files/EHT_memo_Issaoun_2017-CE-03.pdf}
\bibitem[Issaoun et al.(2018)]{issaounetal18} Issaoun S., Falcke, H., Friberg, P., Janssen, M., Tilanus, R., and Wouterloot, J. 2018, EHT memo series, 2018-CE-01 \url{https://eventhorizontelescope.org/files/eht/files/EHT_memo_Issaoun_2018-CE-01.pdf}
\bibitem[Janssen et al.(2019)]{janssenetal19} Janssen, M., Blackburn, L., Issaoun, S., Krichbaum, T. P., Wielgus, M. 2019, EHT memo series, 2019-CE-01 \\ \url{https://eventhorizontelescope.org/files/eht/files/EHT_memo_Janssen_2019-CE-01.pdf}
\bibitem[Koay et al.(2023)]{koayetal23} Koay, J. Y., et al. 2023, EHT memo series, 2023-L1-02, arXiv:2312.02759 [astro-ph.IM]
\bibitem[Kutner \& Ulich(1981)]{kutnerulich81} Kutner M.~L. \& Ulich B.~L., 1981, ApJ, 250, 341
\bibitem[Mart\'{\i}-Vidal et al.(2016)]{martividaletal16} Mart\'{\i}-Vidal I., Roy, A., Conway, J., \& Zensus, A. J. 2016, A\&A, 587, A143
\bibitem[Matsushita et al.(2006)]{matsushitaetal06}Matsushita, S., Saito, M., Sakamoto, K., et al. 2006, in Proceedings of the SPIE, Vol. 6275, Society of Photo-Optical Instru- mentation Engineers (SPIE) Conference Series, 62751W
\bibitem[McMullin et al.(2007)]{mcmullinetal07} McMullin, J. P., Waters, B., Schiebel, D., Young, W., and Golap, K. 2007, in Shaw, R. A., Hill, F., Bell, D. J. eds, ASP Conference Series, Vol. 376, Astronomical Data Analysis Software and Systems XVI. p. 127
\bibitem[Rohlfs \& Wilson(2006)]{rohlfsandwilson06} Rohlfs, K. \& Wilson, T. L. 2006, Tools of Radio Astronomy, Fourth Revised and Enlarged Edition, Corrected Second Printing, Springer
\end{thebibliography}
\end{document}